\documentclass[aps, preprint, superscriptaddress]{revtex4-2}
\usepackage{graphicx}
\usepackage{makecell, multirow}
\usepackage{color, bm, soul}
\usepackage{amsmath, amssymb}
\usepackage{gensymb}
\usepackage{geometry}
\usepackage{braket}
\usepackage{lineno}
\usepackage[colorlinks=true, linkcolor=black, citecolor=black, urlcolor=blue]{hyperref}
\newcommand{\ie}{{\em i.\,e.}}

\newcommand{\etal}{{\em et al.\ }}

\begin{document}


\title{Epitaxial ferroelectric hafnia stabilized by symmetry constraints}

\author{Tianyuan Zhu}
\affiliation{Key Laboratory for Quantum Materials of Zhejiang Province, Department of Physics, School of Science, Westlake University, Hangzhou, Zhejiang 310024, China}
\affiliation{Institute of Natural Sciences, Westlake Institute for Advanced Study, Hangzhou, Zhejiang 310024, China}

\author{Shiqing Deng}
\affiliation{Beijing Advanced Innovation Center for Materials Genome Engineering, University of Science and Technology Beijing, Beijing 100083, China}

\author{Shi Liu}
\email{liushi@westlake.edu.cn}
\affiliation{Key Laboratory for Quantum Materials of Zhejiang Province, Department of Physics, School of Science, Westlake University, Hangzhou, Zhejiang 310024, China}
\affiliation{Institute of Natural Sciences, Westlake Institute for Advanced Study, Hangzhou, Zhejiang 310024, China}

\begin{abstract}{
Ferroelectric memories experienced a revival in the last decade due to the discovery of ferroelectricity in HfO$_2$-based nanometer-thick thin films. These films exhibit exceptional silicon compatibility, overcoming the scaling and integration obstacles that impeded perovskite ferroelectrics' use in high-density integrated circuits. The exact phase responsible for ferroelectricity in hafnia films remains debated with no single factor identified that could stabilize the ferroelectric phase thermodynamically. Here, supported by density functional theory (DFT) high-throughput (HT) calculations that screen a broad range of epitaxial conditions, we demonstrate conclusively that specific epitaxial conditions achievable with common substrates such as yttria-stabilized zirconia (YSZ) and SrTiO$_3$ can favor the polar $Pca2_1$ phase thermodynamically over other polar phases such as $R3m$ and $Pmn2_1$ and nonpolar $P2_1/c$ phase. The substrate's symmetry constraint-induced shear strain is crucial for the preference of $Pca2_1$. The strain-stability phase diagrams resolve experiment-theory discrepancies and can guide the improvement of ferroelectric properties of epitaxial hafnia thin films.
}
\end{abstract}

\maketitle

\clearpage

\section{Introduction}

The fluorite-structured binary oxide, HfO$_2$, is known to form many nonpolar polymorphs including the monoclinic ($M$) $P2_1/c$, tetragonal ($T$) $P4_2/nmc$, and cubic $Fm\overline{3}m$ phases, among which the most stable phase is the $M$ phase~\cite{Ohtaka01p1369}. The observed ferroelectricity in HfO$_2$-based thin films has been attributed to the polar orthorhombic $Pca2_1$ phase \cite{Boscke11p102903, Park15p1811, Shimizu15p032910, Katayama16p134101, Shimizu16p32931, Katayama16p112901, Mimura19pSBBB09, Estandia19p1449, Yun22p903}, while other polar phases such as rhombohedral $R3m$ and $R3$ phases \cite{Wei18p1095, Begon-Lours20p043401, Zheng21p172904} and orthorhombic $Pmn2_1$ phase \cite{Huan14p064111, Qi20p257603} have also been proposed (Fig.~S1). One of the controversies surrounding ferroelectric HfO$_2$ stems from the fact that all these polar phases are higher in energy than the nonpolar $M$ phase (Table~S1). Several extrinsic factors have been suggested to explain the stabilization of the polar phases in thin films. Among them, the surface energy effect has been commonly cited as the primary mechanism that favors $Pca2_1$ and $T$ phases thermodynamically over the $M$ phase in nanocrystals with a high surface-to-volume ratio \cite{Materlik15p134109}.
However, comprehensive surface energy calculations involving multiple major crystallographic orientations revealed that the $M$ phase actually possesses lower surface energy than $Pca2_1$ \cite{Batra16p172902, Wu20p252904}. The impact of defects such as dopants and oxygen vacancy has been considered. DFT studies predicted that even a high concentration of dopants is not enough to reverse the relative stability between the $M$ and polar phases \cite{Batra17p9102, Materlik18p164101}. More recently, it was proposed that charged oxygen vacancies could promote nonpolar-polar phase transitions of HfO$_2$, offering an explanation for the origin of ferroelectricity from the perspective of polymorphism kinetics \cite{Ma23p096801}. We aim to identify a single, readily-tunable parameter that can stabilize the polar phases thermodynamically over the $M$ phase.

Polycrystalline films of hafnia often exhibit a mixture of polar and nonpolar phases, which poses a challenge in isolating the individual contributions of various factors that influence the ferroelectricity. Thin film epitaxy with precisely controlled substrate-ferroelectric interfaces and microstructures serves as an ideal platform to understand the ferroelectric behavior of hafnia \cite{Schroeder22p653}. By leveraging the lattice mismatch between the film and substrate \cite{Paull22p74}, as well as the substrate symmetry and vicinality, it is possible to control the phase stability of hafnia polymorphs for optimal ferroelectric properties and device prototyping. Ferroelectric Y-doped HfO$_2$ (YHO) thin films with various orientations (\{001\}, \{110\} and \{111\}) were grown through lattice-matching epitaxy (LME), each coherently matching the ITO/YSZ substrates (where ITO refers to the indium-tin oxide electrode)~\cite{Shimizu15p032910, Katayama16p134101, Shimizu16p32931, Katayama16p112901, Mimura19pSBBB09}, and the resulting polar phase was identified as $Pca2_1$. However, \citeauthor{Wei18p1095} reported the formation of a compressively strained rhombohedral $R3m$ phase in (111)-oriented Hf$_{0.5}$Zr$_{0.5}$O$_2$ (HZO) thin films deposited on a (001)-oriented (La,Sr)MnO$_3$ (LSMO) electrode and SrTiO$_3$ (STO) substrate~\cite{Wei18p1095}, while another polar rhombohedral $R3$ phase was suggested in HZO thin films grown on the GaN(0001)/Si(111) substrate~\cite{Begon-Lours20p043401}. \citeauthor{Yun22p903} instead demonstrated a rhombohedrally distorted $Pca2_1$ phase in YHO(111) thin films on both LSMO/STO(001) and LSMO/STO(110) substrates~\cite{Yun22p903}. Recently, \citeauthor{Liu23p2209925} suggested both $R3m$ and $Pca2_1$ phases in HZO(111) on LSMO(110)~\cite{Liu23p2209925}. Because of the large lattice mismatch between LSMO and HZO, HZO films grown on LSMO/STO likely adopt the domain-matching epitaxy (DME) where $m$ lattices of film match $n$ lattices of substrate (Fig.~S2 and Table~S2). 

Several DFT studies have attempted to reveal the impact of epitaxial strains on the relative stability of various hafnia polymorphs, but the findings have been inconsistent.  \citeauthor{Qi20p257603} proposed that an in-plane shear strain could promote $T\rightarrow Pmn2_1$ transition and attributed the ferroelectricity of HZO(111) on LSMO/STO(001) to the kinetically stabilized $Pmn2_1$ phase~\cite{Qi20p257603}. \citeauthor{Zheng21p172904} suggested that the trigonal symmetry constraint imposed by the ZnO(0001) substrate renders the $R3m$ phase energetically competitive with the $M$ phase~\cite{Zheng21p172904}, although the possibility of a distorted $Pca2_1$ phase can not be ruled out.
Furthermore, different studies have argued that compressive~\cite{Liu19p054404} and tensile strains~\cite{Chae20p054101} may be responsible for stabilizing the $Pca2_1$ phase. To summarize, there is currently no consensus either experimentally or theoretically on the following key questions: (i) which polar phase or phases ($Pca2_1$, $Pmn2_1$, $R3m$, and $R3$) are responsible for the ferroelectricity in epitaxial thin films? (ii) what types of strains, tensile or compressive, can stabilize the polar phase? (iii) can a single factor stabilize the ferroelectric phase thermodynamically in hafnia thin films?

In this work, we address the aforementioned questions by performing DFT-based high-throughput (HT) calculations on $\approx$3500 configurations to quantitatively assess the influence of a broad range of isotropic and anisotropic epitaxial strains, as well as substrate symmetry, on the phase competitions in thin films of HfO$_2$. We show that instead of focusing on the type of stain, either compressive or tensile, applied to the ground state of a polymorph (as commonly done in prior studies), a more conceptually straightforward and experimentally relevant approach is to examine the influence of a given substrate on the relative phase stability in the film.  
Our results  provide definitive proof that the $Pca2_1$ phase can be intentionally engineered as the most stable phase across an extensive range of epitaxial conditions that impose orthogonal in-plane lattices, thereby resolving multiple discrepancies between experimental and theoretical observations.

\section{Results and discussion}

\begin{figure}[t]
\includegraphics[width=3.4 in]{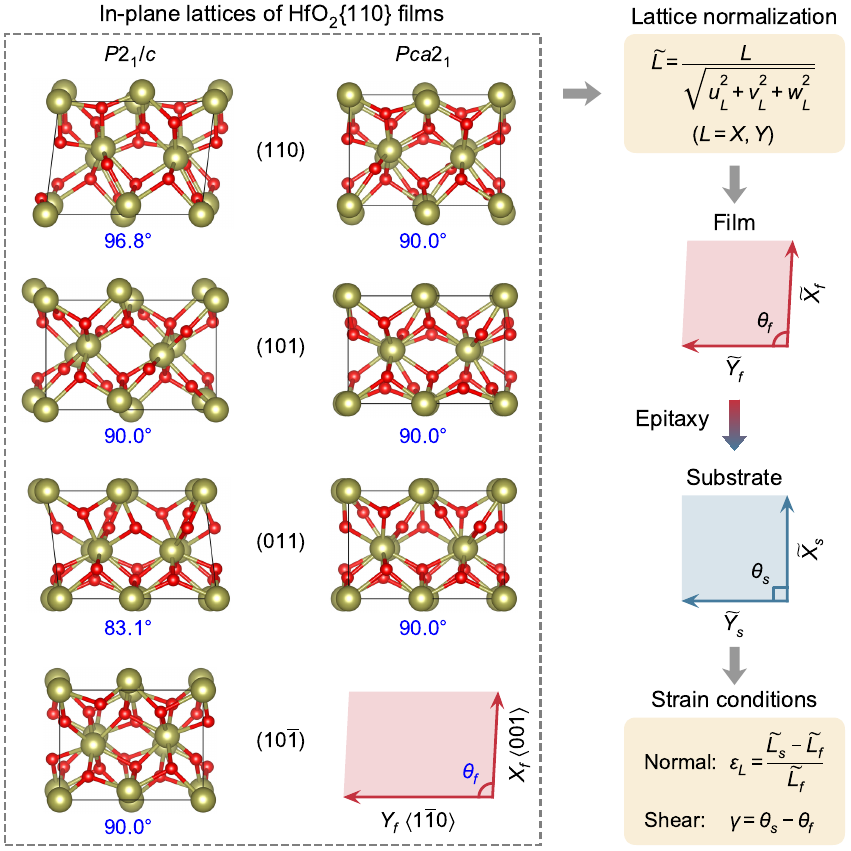}
\caption{\textbf{Epitaxial matching of different HfO$\bm{_2}$$\bm{\{110\}}$ films with a generic substrate and resulted strain conditions.} The left panel shows the in-plane lattices for four $P2_1/c$ and three $Pca2_1$ unique growth orientations, each of which is characterized by a set of lattice parameters ($X_f$, $Y_f$, $\theta_f$). By normalizing the in-plane lattice parameters as ($\widetilde{X}_f$, $\widetilde{Y}_f$, $\theta_f$), the strain conditions ($\varepsilon_X$, $\varepsilon_Y$, $\gamma$) of these films imposed by a generic substrate ($\widetilde{X}_s$, $\widetilde{Y}_s$, $\theta_s$) depend on both growth orientation and crystal symmetry.}
\label{fig_symmetry}
\end{figure}

We start by emphasizing that the epitaxial constraints experienced by a film grown on a given substrate depends on the film-substrate matching plane ($\sigma$) represented by Miller indices $(hkl)$ and the crystal symmetry ($\varphi$) of the polymorph (see discussions below). $\sigma$ labels the growth orientation while $\varphi$ determines the number of unique growth orientations within the same family of $\{hkl\}$ (referred to as ``general orientation"). As shown in Fig.~\ref{fig_symmetry}, in the case of \{110\}-oriented thin films, there exist four unique growth orientations for $M$ but only three for $Pca2_1$. For convenience, we introduce the method of lattice normalization for polymorphs in films ($f$), $\widetilde{L}_f = L_f/\sqrt{u_L^2+v_L^2+w_L^2}$, where $L_f$ ($L=X, Y$) is the length of an in-plane lattice vector $[u_Lv_Lw_L]\in(hkl)$ that is aligned along the measurement axis $X$ or $Y$ (Table~S3), and the intrinsic lattice angle is denoted as $\theta_f$. Similarly, the mechanical boundary conditions of a {\em generic} substrate ($s$) can be specified with normalized in-plane lattice constants ($\widetilde{X}_s$ and $\widetilde{Y}_s$) and lattice angle $\theta_s$. 

\begin{figure}[t]
\includegraphics[width=5.6 in]{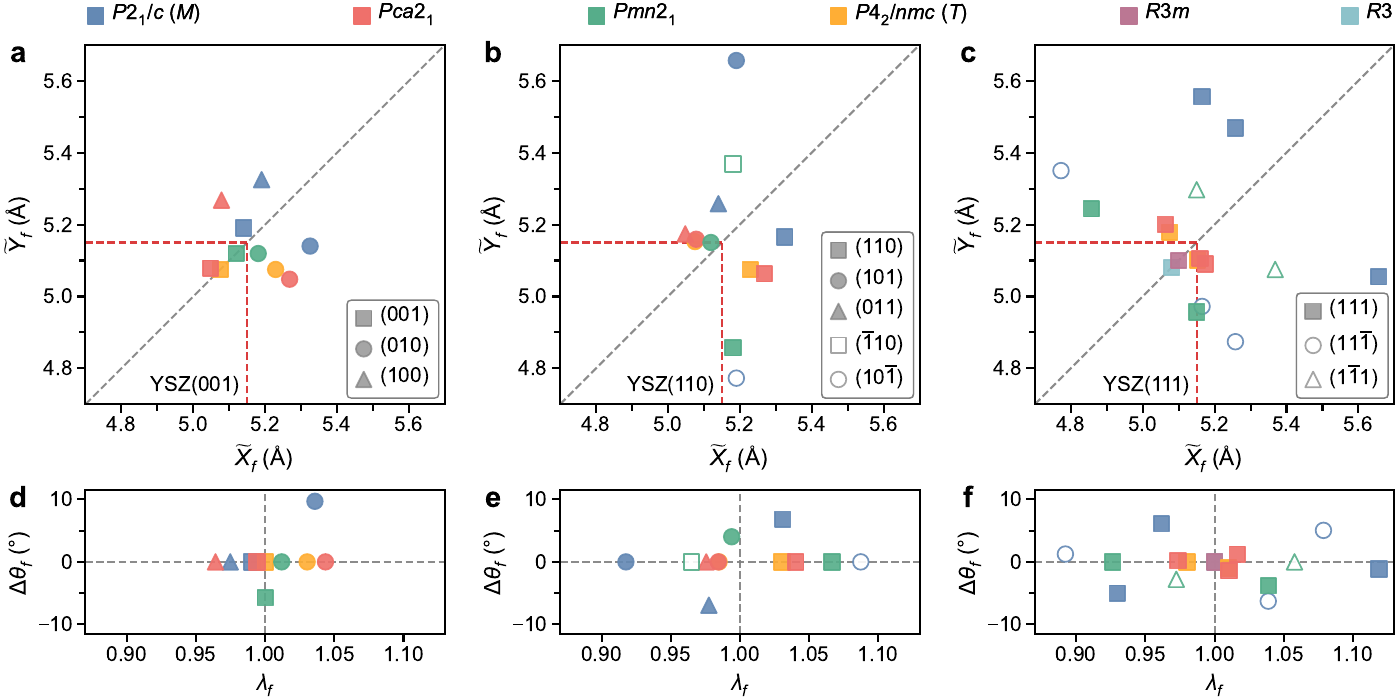}
\caption{\textbf{Ground-state epitaxial conditions of unstrained HfO$_2$ polymorphs of different growth orientations.} The top panels show the normalized lattice lengths ($\widetilde{X}_f$, $\widetilde{Y}_f$) of (\textbf{a}) \{001\}, (\textbf{b}) \{110\}, and (\textbf{c}) \{111\}-oriented polymorphs. Different phases and growth orientations are denoted by different colored and shaped markers, respectively. Note that for the strain-induced polar rhombohedral $R3m$ phase, the unstrained state is considered as the nonpolar cubic $P\overline{4}3m$ phase \cite{Schroeder22p653} (Table S1). The red dashed lines mark the experimental lattice constant (5.15 \AA) of the YSZ substrate \cite{Torrejon18p013401}. The bottom panels show the distortion angle $\Delta \theta_f=\theta_f-90\degree$ and the anisotropic parameter $\lambda_f = \widetilde{X}_f/\widetilde{Y}_f$ of (\textbf{d}) \{001\}, (\textbf{e}) \{110\}, and (\textbf{f}) \{111\}-oriented polymorphs. The grey dashed lines in all panels denote the epitaxial conditions of isotropic substrates with $\widetilde{X}_s = \widetilde{Y}_s$ and $\theta_s=90\degree$.}
\label{fig_epitaxy}
\end{figure}

A key aspect of hafnia epitaxy, which is often underappreciated, is that a given substrate ($\widetilde{X}_s$, $\widetilde{Y}_s$, $\theta_s$) can impose drastically different strain conditions depending on the values of $\sigma$ and $\varphi$ associated with the crystallized polymorph in the film.
This can be understood by examining the ground-state epitaxial conditions of unstrained hafnia polymorphs.
We define an anisotropic parameter,  $\lambda_f = \widetilde{X}_f/\widetilde{Y}_f$, and a distortion angle, $\Delta \theta_{f} = \theta_f - \theta_s$ (with $\theta_s=90 \degree$ for simplicity). The degree of in-plane anisotropy for a polymorph is quantified by the deviation of $\lambda_f$ from unity and the amount of in-plane shear strain experienced by the film scales with the value of $\Delta \theta_{f}$. The four parameters, $\widetilde{X}_f$, $\widetilde{Y}_f$, $\lambda_f$, and $\Delta \theta_f$, hence collectively characterize the strain-free epitaxial condition for a specific phase and growth orientation. Figure~\ref{fig_epitaxy} plots the values of ground-state epitaxial conditions for a number of phases of HfO$_2$ with $\{001\}$, $\{110\}$, and $\{111\}$ orientations, respectively, revealing several important characteristics. First, using the HfO$_2$/YSZ heterostructure grown by LME as an example, the normal strain, $\varepsilon_L = (\widetilde{L}_s-\widetilde{L}_f)/\widetilde{L}_f$, clearly depends on both $\sigma$ and $\varphi$ of the polymorph. Specifically, YSZ ($a_{\mathrm{YSZ}} = 5.15$ \AA) with identical \{110\} orientations induces a non-equibiaxial compressive strain in the (101)-oriented $M$ phase but results in $\varepsilon_X=-0.8\%$ and $\varepsilon_Y=+7.9\%$ for the ($10\overline{1}$)-oriented $M$ phase. Second, the $T$ phase and $Pca2_1$ phase of the same orientation always have similar ground-state epitaxial conditions that are close to be isotropic as characterized by $\lambda_f=1$ and $\Delta \theta_f =0$. In contrast, the values of $\lambda_f$ and $\Delta \theta_f$ for $M$ and $Pmn2_1$ (blue and green markers in Fig.~\ref{fig_epitaxy}) deviate more from the isotropic condition.  Finally, for $\{111\}$-oriented films, the ground-state epitaxial conditions of $Pca2_1$, $T$, $R3m$, and $R3$ phases are comparably close to those of the isotropic YSZ substrate. This could make it challenging to distinguish between these phases in experimental settings.

Because the same substrate can cause varying strains in the film, it is essential to thoroughly consider all possible growth orientations of competing polymorphs to establish the correct thermodynamic stability order. 
Our HT DFT calculations are performed by computing phase energetics for an expansive range of ($\widetilde{X}_s$, $\widetilde{Y}_s$) values with $\theta_s=90\degree$ for three general orientations: $\{001\}$, $\{110\}$, and $\{111\}$, respectively. Given a substrate and general orientation, we optimize the supercell structures for all possible $\sigma$ and $\varphi$ values. During the structural optimization, the in-plane lattice parameters remain fixed to those of the substrate ($\widetilde{X}_s$, $\widetilde{Y}_s$) while the atomic coordinates and out-of-plane lattice parameters are allowed to relax (see Methods).

\begin{figure}[t]
\includegraphics[width=3.2 in]{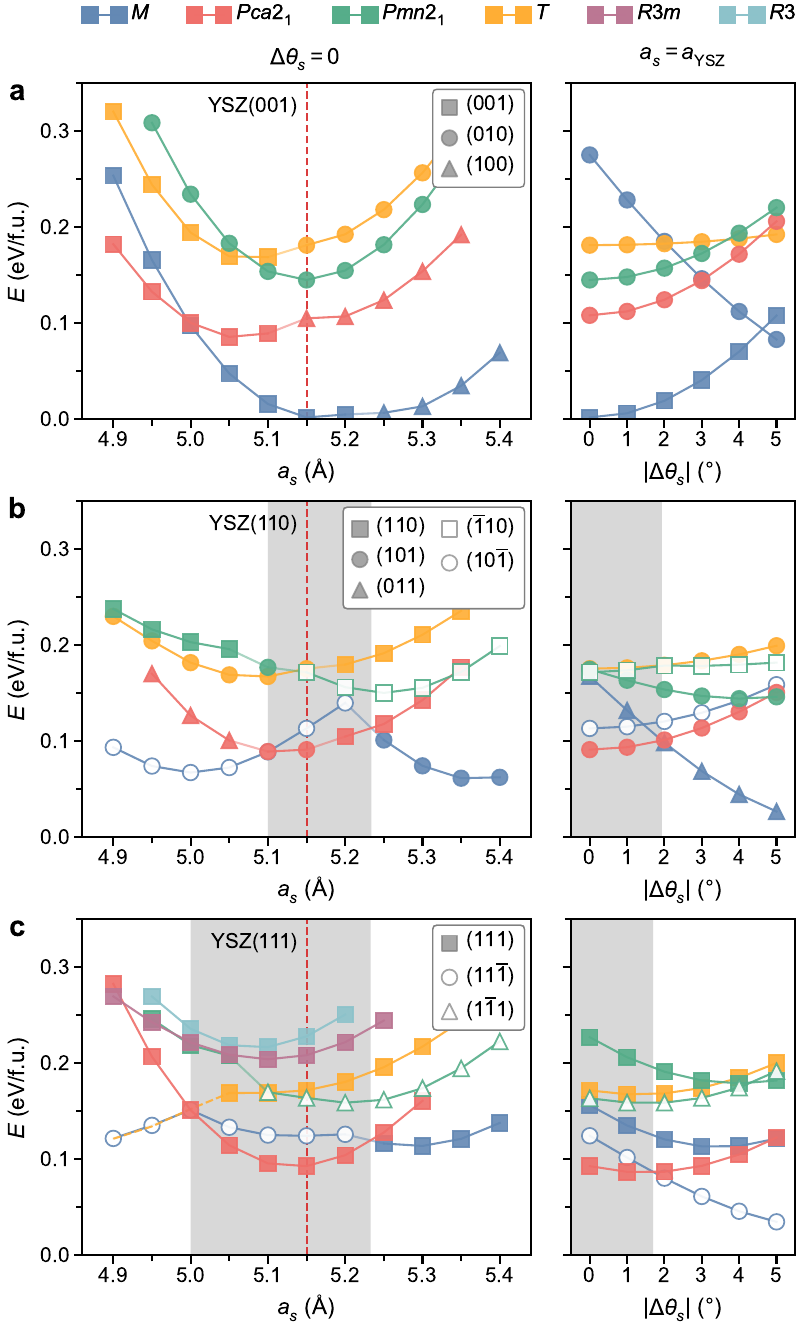}
\caption{\textbf{Thermodynamic stability of HfO$\bm{_2}$ thin films under isotropic epitaxial conditions.} The left panels show the energy of the most stable orientation of a given phase as a function of the substrate lattice constant $a_s$ in (\textbf{a}) \{001\}, (\textbf{b}) \{110\}, and (\textbf{c}) \{111\}-oriented films. Different phases and growth orientations are denoted by different colored and shaped markers, respectively. The red dashed lines mark the YSZ lattice constant ($a_{\mathrm{YSZ}}=5.15$ \AA). The right panels display the energy as a function of the substrate distortion angle $\left|\Delta \theta_s \right|= \left|\theta_s-90\degree \right|$ when $a_s$ is fixed as $a_{\mathrm{YSZ}}$. The grey shaded regions mark the ranges of epitaxial conditions where the polar $Pca2_1$ phase is the most stable phase. The energy of the unstrained $M$ phase is set to zero as reference.}
\label{fig_isotropy}
\end{figure}

We first investigate the effects of isotropic epitaxial constraints ($\widetilde{X}_s=\widetilde{Y}_s=a_s$ and $\theta_s=90\degree$) on the phase competitions in differently-oriented hafnia thin films. Note that most of previous theoretical studies only considered equibiaxial strains, which involve equal scaling lattice vectors along $X$ and $Y$ while conserving the in-plane lattice angle at the ground-state value of $\theta_f$. That is, the in-plane shear strain is not considered as $\Delta \theta_f=0$. Here, closely resembling LME on isotropic substrates such as YSZ, each hafnia polymorph has in-plane lattice vectors being orthogonal and the lengths fixed to $a_s$. Depending on the values of $\sigma$ and $\varphi$, a hafnia polymorph could be subjected to in-plane shear strain characterized by $\Delta \theta_f$ as discussed above. The value of $a_s$ varies from 4.9 to 5.4~\AA.
Figure~\ref{fig_isotropy} shows the calculated phase energetics for $M$, $Pca2_1$, $Pmn2_1$, $T$, $R3m$, and $R3$ with different growth orientations as functions of $a_s$. For a specific $a_s$, only the lowest energy value of each phase is plotted.

In the case of \{001\}-oriented polymorphs (Fig.~\ref{fig_isotropy}\textbf{a}), the most stable phase on  YSZ(001) is the $M$ phase with the same orientation, consistent with the report by Torrejon~\etal~which demonstrated the formation of a slightly distorted $M$ phase with equal in-plane lattice constants in HZO/YSZ(001)~\cite{Torrejon18p013401}.
This is expected because the strain-free epitaxial condition of $M$(001) is close to that of YSZ(001) as shown in Fig.~\ref{fig_epitaxy}\textbf{a}.
In addition, a critical value of $a_s=5.0$~\AA~is identified below which $Pca2_1$(001) becomes most stable. However, this film will only exhibit in-plane polarization ($Pca2_1$ has polarization along [010]) that is not convenient for device lateral downscaling. Overall, $M$(001) is the most stable polymorph over a wide range of isotropic epitaxial conditions.

For \{110\}-oriented polymorphs (Fig.~\ref{fig_isotropy}\textbf{b}), $Pca2_1$ becomes more stable than $M$ within a specific strain range of $5.10<a_s< 5.23$~\AA, and the energies of two orientations of $Pca2_1$, (101) and (011), are almost equal, indicating a degenerate energy landscape. This is supported by experimentally observed coexistence of $Pca2_1$(101) and (011) domains in YHO/YSZ(110)~\cite{Shimizu16p32931}. Specifically, when grown on YSZ(110), our calculations indicate that the energy of $Pca2_1$ is 22 meV per formula unit (f.u.) lower than $M$, and a larger $a_s=5.20$~\AA~can further increase this energy difference to 35 meV/f.u. Outside the strain range that stabilize $Pca2_1$, a larger $a_s$ favors the formation of $M$(101), while a smaller $a_s$ promotes $M$($10\overline{1}$). 

\begin{figure}[t]
\includegraphics[width=4.5 in]{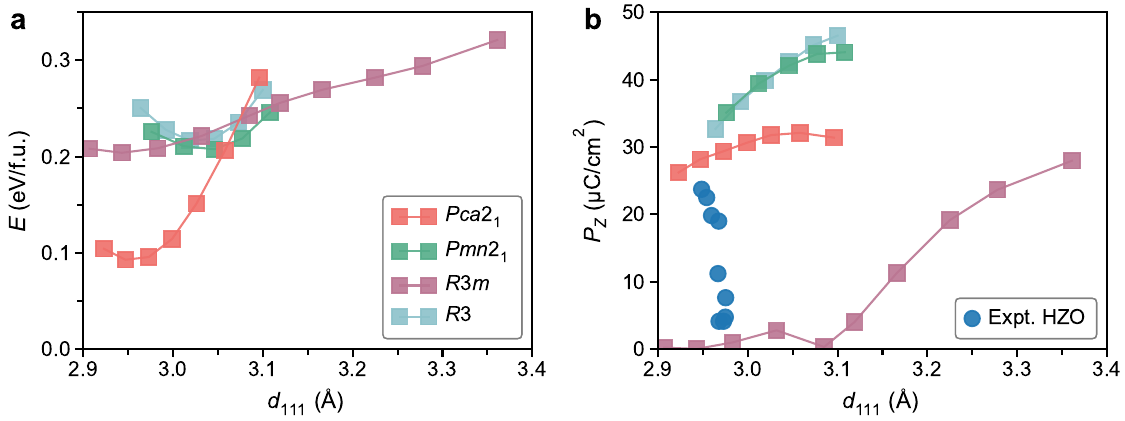}
\caption{\textbf{Energy and polarization of four polar phases in HfO$\bm{_2}$(111) thin films under isotropic epitaxial conditions.} (\textbf{a}) Energy and (\textbf{b}) out-of-plane polarization $P_Z$ as a function of interplanar spacing $d_{111}$. The blue circles mark the experimental remanent polarization and $d_{111}$ of HZO epitaxial thin films grown on different substrates \cite{Estandia19p1449}.}
\label{fig_polarization}
\end{figure}

Regarding \{111\}-oriented polymorphs (Fig.~\ref{fig_isotropy}\textbf{c}), the $Pca2_1$ phase can be stabilized by isotropic substrates with $a_s$ in a wide range of 5.00--5.23~\AA, outside which the $M$ phase is more stable. Importantly, we find that the energies of polar $R3m$ and $R3$ phases are  considerably higher than the other phases. Although $R3m$ becomes competitive with $Pca2_1$ under large compressive strains ($a_s <4.9$ \AA), the nonpolar $M$ phase remains the most stable at these epitaxial conditions. As pointed out by \citeauthor{Fina21p1530} \cite{Fina21p1530}, the assumption of the $R3m$ phase in (111)-oriented HZO films leads to an important mismatch between DFT calculations and experiments on the required strain for the measured polarization. The strain state can be related to the out-of-plane interplanar spacing $d_{111}$ in (111)-oriented films. We calculate the energy and polarization of four polar phases ($Pca2_1$, $Pmn2_1$, $R3m$, and $R3$) as a function of $d_{111}$, respectively (Fig.~\ref{fig_polarization}). It is found that when $d_{111}<3.06$ \AA, $Pca2_1$ is most stable with an out-of-plane polarization of $\sim$30 $\mu$C/cm$^2$, comparable to experimental values of $\approx$4--23 $\mu$C/cm$^2$ considering the depolarization effect. In contrast, a giant value of $d_{111}$ of  3.4 \AA~ in the $R3m$ phase is needed to induce a polarization of the same magnitude. Since the experimental values of $d_{111}$ fall within the range of 2.96--3.05~\AA~as reported in a few HZO(111) films grown on different substrates, we believe $Pca2_1$ is responsible for the ferroelectricity in (111)-oriented epitaxial hafnia thin films.

Our extensive investigations on all growth orientations in the family of $\{001\}$, $\{110\}$, and $\{111\}$ demonstrate that the epitaxial strain can serve as the sole factor that thermodynamically stabilizes $Pca2_1$ over $M$, offering a straightforward explanation to the origin of ferroelectricity in epitaxial HfO$_2$-based thin films on YSZ substrates reported in experiments. This strain effect has been elusive because previous DFT studies either overlooked certain low-energy orientations~\cite{Liu19p054404} or disregarded the in-plane shear strain that results from the orthogonal lattice vectors of isotropic substrates~\cite{Zhang20p014068}. We now prove that the in-plane shear strain is crucial for the stabilization of $Pca2_1$ by performing a series of model calculations that estimate the phase energetics on (hypothetical) substrates with $a_s$ fixed to YSZ lattice constant but varying $\theta_s$. This enables the isolation of the shear strain contribution, \ie, the (110)-oriented $M$ phase ($\theta_f=96.8\degree$) grown on a substrate with $\theta_s=92\degree$ has $\Delta \theta_f=4.8\degree$, thereby experiencing a smaller shear strain compared to that on a substrate of $\theta_s=90\degree$. 

The energies of representative polymorphs as a function of $|\Delta \theta_s|=|\theta_s-90\degree|$ are presented in the right panels of Fig.~\ref{fig_isotropy}. In \{001\}-oriented films, most polymorphs show increasing energy with increasing $|\Delta \theta_s|$ as their values of $\theta_f$ are already close to 90$\degree$, matching well to a substrate of $\theta_s=90\degree$. For $\Delta \theta_s<3.0\degree$, $Pca2_1$(010) is lower in energy than $M$(010) but still higher in energy than $M$(001). It may be feasible to obtain ferroelectric (010)-oriented films on YSZ by finding a way to prevent the formation of $M$(001). For \{110\}-oriented polymorphs, when $\Delta \theta_s<1.9\degree$, $Pca2_1$ is preferred over $M$. Furthermore, as $\Delta \theta_s$ increases and substrate $\theta_s$ approaches the value of $\theta_f=83.1\degree$ of $M$(011), the energy of $M$(011) reduces considerably. This implies that in the absence of shear strain applied to $M$(011), the nonpolar $M$ phase will be highly favored over $Pca2_1$, as is the case in bulk, underscoring the substantial impact of shear strain on the stability of the $M$ phase. Another finding is that at $\Delta \theta_s=0\degree$, $M(10\overline{1})$ having $\theta_f=90\degree$ is higher in energy than $Pca2_1$, mainly due to its large strain-free anisotropy ($\lambda_f=1.09$). Regarding \{111\} orientations, $Pca2_1$ is most stable when $\Delta \theta_s$ values are below 1.7$\degree$. Decreasing shear strain by increasing $\Delta \theta_s$ leads to a significant reduction in energy for $M(11\overline{1})$. Our results also reveal that on a substrate of $\theta_s=90\degree$, the $T$ phase is more stable than $Pmn2_1$, challenging the hypothesis of spontaneous transition of $T\rightarrow Pmn2_1$~\cite{Qi20p257603}.

\begin{figure}[t]
\includegraphics[width=6.8 in]{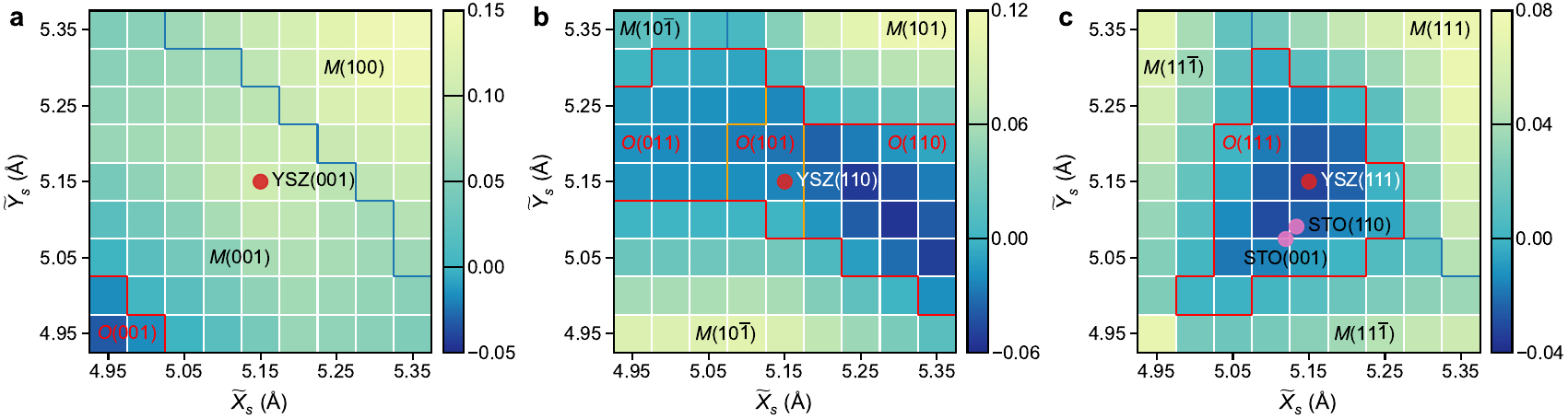}
\caption{\textbf{Strain-stability phase diagrams of HfO$\bm{_2}$ thin films.} The color scales with the energy difference (in unit of eV/f.u.) between the most stable polar $Pca2_1$ phase (labeled as $O$) and the most stable nonpolar $M$ phase for (\textbf{a}) \{001\}, (\textbf{b}) \{110\}, and (\textbf{c}) \{111\}-oriented films. The red lines denote the phase boundaries between $Pca2_1$ and $M$ phases, while the blue and orange lines separate the $M$ phase and $Pca2_1$ phase of different orientations, respectively. Experimental epitaxial conditions (Table S2) of the isotropic YSZ \cite{Torrejon18p013401} and anisotropic STO \cite{Yun22p903} substrates are marked by red and pink circles, respectively.}
\label{fig_anisotropy}
\end{figure}

For substrates like LSMO/STO that have a large lattice mismatch with HfO$_2$, DME results in smaller effective mismatch than LME~\cite{Estandia20p3801, Jiao21p252901}, and the effective strain experienced by the film could be anisotropic. In this regard, we further study the phase competitions under anisotropic epitaxial conditions characterized by $\widetilde{X}_s \neq \widetilde{Y}_s$ and $\theta_s=90\degree$. As demonstrated in Fig.~\ref{fig_isotropy} that $M$ and $Pca2_1$ consistently have lower energies than the other phases, here we focus on these two phases. Figure~\ref{fig_anisotropy} displays the phase diagrams of HfO$_2$ thin films under anisotropic epitaxial conditions, with color representing the energy difference ($\Delta E$) between the most stable $Pca2_1$ phase and the most stable $M$ phase for a given general orientation. In \{001\}-oriented films, anisotropic epitaxial conditions accessible in experiments strongly promote the formation of $M$ in both (001) and (100) orientations, similar to the isotropic case presented in Fig.~\ref{fig_isotropy}\textbf{a}. For $\{110\}$ orientations, the energy difference $\Delta E$ is typically more responsive to the strain applied along the $Y\braket{1\overline{1}0}$ direction (Fig.~\ref{fig_symmetry}). A uniaxial tensile strain along $X$ can further facilitate the formation of $Pca2_1$(110). In $\{111\}$ orientations, the $Pca2_1$ phase is stabilized when the normalized lattice lengths $\widetilde{X}_s$ and $\widetilde{Y}_s$ range from 5.05 to 5.25 \AA. Notably, both isotropic YSZ(111) and anisotropic STO(001)/(110) epitaxial conditions fall within this range (Fig.~\ref{fig_anisotropy}\textbf{c}). Moreover, the effect of shear strain combined with anisotropic normal strain on the phase stability is further examined. We map out the phase diagrams for \{110\}- and \{111\}-oriented polymorphs with $\Delta \theta_s=1.5\degree$ and  $\Delta \theta_s=1.0\degree$, respectively (Fig.~S3). We find that a substrate featuring non-orthogonal in-plane lattice vectors generally constricts the range of epitaxial conditions that can stabilize $Pca2_1$, due to the diminished shear strain applied to $M$.

The strain-stability phase diagrams established with HT DFT calculations provide answers to the three questions raised above. First, our findings indicate that the $Pca2_1$ phase is most likely the ferroelectric phase formed in epitaxial hafnia thin films grown by LME and DME. Other polar phases such as $R3m$, $R3$, and $Pmn2_1$ are all higher in energy than $Pca2_1$ across a wide range of epitaxial conditions. The experimentally observed rhombohedral symmetry~\cite{Wei18p1095} could stem from the lattice distortion of $Pca2_1$ phase. Second, as the same substrate can generate varying strain conditions, classifying the strain type, be it tensile or compressive, that stabilizes the polar phase is not particularly useful. Instead, we recommend focusing on the effective epitaxial conditions of the substrate. Finally, the $Pca2_1$ phase can be stabilized across a broad range of epitaxial conditions in both $\{110\}$ and $\{111\}$ growth orientations by imposing orthogonal lattice constraints using substrates with orthogonal in-plane lattice vectors. These epitaxial conditions primarily destabilize the $M$ phase with a large intrinsic in-plane lattice angle and/or significant anisotropic ratio. Moreover, our results offer a potential explanation for the \textit{reverse size effect} observed in ferroelectric HfO$_2$-based thin films \cite{Cheema20p478}: with the film thickness increasing, the relaxation of epitaxial constraints, particularly the shear strain, restores the thermodynamic stability of $M$, leading to a suppressed ferroelectricity in thicker films.

In summary, this study demonstrates that the epitaxial conditions presented in common substrates such as YSZ and STO can thermodynamically stabilize \{110\}- and (111)-oriented polar $Pca2_1$ phase without relying on other extrinsic factors. The shear strain arising from the symmetry of the substrate that tends to orthogonalize in-plane lattices plays a crucial role in destabilizing the nonpolar $M$ phase. By clarifying the ambiguities surrounding the field of ferroelectric hafnia, we hope to facilitate the optimization of epitaxial hafnia thin films, ultimately leading to enhanced functionalities. 

\section{Methods}

DFT calculations are performed using the Vienna \textit{ab initio} simulation package (VASP) \cite{Kresse96p11169} with the projector augmented-wave (PAW) method \cite{Blochl94p17953, Kresse99p1758} and the Perdew-Burke-Ernzerhof (PBE) exchange correlation functional \cite{Perdew96p3865}. The plane-wave cutoff energy is set to 600 eV. The Brillouin zones of the \{001\}, \{110\}, and \{111\} supercells are sampled by $\Gamma$-centered (4$\times$4$\times$4), (4$\times$3$\times$3), and (3$\times$2$\times$3) Monkhorst-Pack \cite{Monkhorst76p5188} $k$-point meshes, respectively. For a specific epitaxial condition, the initial structures are constructed by accordingly setting the in-plane lattice parameters, then the atomic coordinates and out-of-plane lattices are fully optimized with a force convergence threshold of 0.01 eV/\AA~with fixed in-plane lattices. The polarization values are calculated by using the Berry phase method \cite{King-Smith93p1651, Vanderbilt93p4442}.

\section{Acknowledgments}

T.Z. and S.L. acknowledge the supports from National Key R\&D Program of China (2021YFA 1202100), National Natural Science Foundation of China (12074319), and Westlake Education Foundation. The computational resource is provided by Westlake HPC Center.





\clearpage

\clearpage

\clearpage

\clearpage

\clearpage

\clearpage

\bibliography{main}

\begin{thebibliography}{39}%
\makeatletter
\providecommand \@ifxundefined [1]{%
 \@ifx{#1\undefined}
}%
\providecommand \@ifnum [1]{%
 \ifnum #1\expandafter \@firstoftwo
 \else \expandafter \@secondoftwo
 \fi
}%
\providecommand \@ifx [1]{%
 \ifx #1\expandafter \@firstoftwo
 \else \expandafter \@secondoftwo
 \fi
}%
\providecommand \natexlab [1]{#1}%
\providecommand \enquote  [1]{``#1''}%
\providecommand \bibnamefont  [1]{#1}%
\providecommand \bibfnamefont [1]{#1}%
\providecommand \citenamefont [1]{#1}%
\providecommand \href@noop [0]{\@secondoftwo}%
\providecommand \href [0]{\begingroup \@sanitize@url \@href}%
\providecommand \@href[1]{\@@startlink{#1}\@@href}%
\providecommand \@@href[1]{\endgroup#1\@@endlink}%
\providecommand \@sanitize@url [0]{\catcode `\\12\catcode `\$12\catcode
  `\&12\catcode `\#12\catcode `\^12\catcode `\_12\catcode `\%12\relax}%
\providecommand \@@startlink[1]{}%
\providecommand \@@endlink[0]{}%
\providecommand \url  [0]{\begingroup\@sanitize@url \@url }%
\providecommand \@url [1]{\endgroup\@href {#1}{\urlprefix }}%
\providecommand \urlprefix  [0]{URL }%
\providecommand \Eprint [0]{\href }%
\providecommand \doibase [0]{https://doi.org/}%
\providecommand \selectlanguage [0]{\@gobble}%
\providecommand \bibinfo  [0]{\@secondoftwo}%
\providecommand \bibfield  [0]{\@secondoftwo}%
\providecommand \translation [1]{[#1]}%
\providecommand \BibitemOpen [0]{}%
\providecommand \bibitemStop [0]{}%
\providecommand \bibitemNoStop [0]{.\EOS\space}%
\providecommand \EOS [0]{\spacefactor3000\relax}%
\providecommand \BibitemShut  [1]{\csname bibitem#1\endcsname}%
\let\auto@bib@innerbib\@empty
\bibitem [{\citenamefont {Ohtaka}\ \emph {et~al.}(2001)\citenamefont {Ohtaka},
  \citenamefont {Fukui}, \citenamefont {Kunisada}, \citenamefont {Fujisawa},
  \citenamefont {Funakoshi}, \citenamefont {Utsumi}, \citenamefont {Irifune},
  \citenamefont {Kuroda},\ and\ \citenamefont {Kikegawa}}]{Ohtaka01p1369}%
  \BibitemOpen
  \bibfield  {author} {\bibinfo {author} {\bibfnamefont {O.}~\bibnamefont
  {Ohtaka}}, \bibinfo {author} {\bibfnamefont {H.}~\bibnamefont {Fukui}},
  \bibinfo {author} {\bibfnamefont {T.}~\bibnamefont {Kunisada}}, \bibinfo
  {author} {\bibfnamefont {T.}~\bibnamefont {Fujisawa}}, \bibinfo {author}
  {\bibfnamefont {K.}~\bibnamefont {Funakoshi}}, \bibinfo {author}
  {\bibfnamefont {W.}~\bibnamefont {Utsumi}}, \bibinfo {author} {\bibfnamefont
  {T.}~\bibnamefont {Irifune}}, \bibinfo {author} {\bibfnamefont
  {K.}~\bibnamefont {Kuroda}},\ and\ \bibinfo {author} {\bibfnamefont
  {T.}~\bibnamefont {Kikegawa}},\ }\bibfield  {title} {\bibinfo {title} {Phase
  relations and volume changes of hafnia under high pressure and high
  temperature},\ }\href {https://doi.org/10.1111/j.1151-2916.2001.tb00843.x}
  {\bibfield  {journal} {\bibinfo  {journal} {J. Am. Ceram. Soc.}\ }\textbf
  {\bibinfo {volume} {84}},\ \bibinfo {pages} {1369} (\bibinfo {year}
  {2001})}\BibitemShut {NoStop}%
\bibitem [{\citenamefont {B{\"o}scke}\ \emph {et~al.}(2011)\citenamefont
  {B{\"o}scke}, \citenamefont {M{\"u}ller}, \citenamefont {Br{\"a}uhaus},
  \citenamefont {Schr{\"o}der},\ and\ \citenamefont
  {B{\"o}ttger}}]{Boscke11p102903}%
  \BibitemOpen
  \bibfield  {author} {\bibinfo {author} {\bibfnamefont {T.~S.}\ \bibnamefont
  {B{\"o}scke}}, \bibinfo {author} {\bibfnamefont {J.}~\bibnamefont
  {M{\"u}ller}}, \bibinfo {author} {\bibfnamefont {D.}~\bibnamefont
  {Br{\"a}uhaus}}, \bibinfo {author} {\bibfnamefont {U.}~\bibnamefont
  {Schr{\"o}der}},\ and\ \bibinfo {author} {\bibfnamefont {U.}~\bibnamefont
  {B{\"o}ttger}},\ }\bibfield  {title} {\bibinfo {title} {Ferroelectricity in
  hafnium oxide thin films},\ }\href {https://doi.org/10.1063/1.3634052}
  {\bibfield  {journal} {\bibinfo  {journal} {Appl. Phys. Lett.}\ }\textbf
  {\bibinfo {volume} {99}},\ \bibinfo {pages} {102903} (\bibinfo {year}
  {2011})}\BibitemShut {NoStop}%
\bibitem [{\citenamefont {Park}\ \emph {et~al.}(2015)\citenamefont {Park},
  \citenamefont {Lee}, \citenamefont {Kim}, \citenamefont {Kim}, \citenamefont
  {Moon}, \citenamefont {Kim}, \citenamefont {M{\"u}ller}, \citenamefont
  {Kersch}, \citenamefont {Schroeder}, \citenamefont {Mikolajick},\ and\
  \citenamefont {Hwang}}]{Park15p1811}%
  \BibitemOpen
  \bibfield  {author} {\bibinfo {author} {\bibfnamefont {M.~H.}\ \bibnamefont
  {Park}}, \bibinfo {author} {\bibfnamefont {Y.~H.}\ \bibnamefont {Lee}},
  \bibinfo {author} {\bibfnamefont {H.~J.}\ \bibnamefont {Kim}}, \bibinfo
  {author} {\bibfnamefont {Y.~J.}\ \bibnamefont {Kim}}, \bibinfo {author}
  {\bibfnamefont {T.}~\bibnamefont {Moon}}, \bibinfo {author} {\bibfnamefont
  {K.~D.}\ \bibnamefont {Kim}}, \bibinfo {author} {\bibfnamefont
  {J.}~\bibnamefont {M{\"u}ller}}, \bibinfo {author} {\bibfnamefont
  {A.}~\bibnamefont {Kersch}}, \bibinfo {author} {\bibfnamefont
  {U.}~\bibnamefont {Schroeder}}, \bibinfo {author} {\bibfnamefont
  {T.}~\bibnamefont {Mikolajick}},\ and\ \bibinfo {author} {\bibfnamefont
  {C.~S.}\ \bibnamefont {Hwang}},\ }\bibfield  {title} {\bibinfo {title}
  {Ferroelectricity and antiferroelectricity of doped thin {HfO$_2$}-based
  films},\ }\href {https://doi.org/10.1002/adma.201404531} {\bibfield
  {journal} {\bibinfo  {journal} {Adv. Mater.}\ }\textbf {\bibinfo {volume}
  {27}},\ \bibinfo {pages} {1811} (\bibinfo {year} {2015})}\BibitemShut
  {NoStop}%
\bibitem [{\citenamefont {Shimizu}\ \emph {et~al.}(2015)\citenamefont
  {Shimizu}, \citenamefont {Katayama}, \citenamefont {Kiguchi}, \citenamefont
  {Akama}, \citenamefont {Konno},\ and\ \citenamefont
  {Funakubo}}]{Shimizu15p032910}%
  \BibitemOpen
  \bibfield  {author} {\bibinfo {author} {\bibfnamefont {T.}~\bibnamefont
  {Shimizu}}, \bibinfo {author} {\bibfnamefont {K.}~\bibnamefont {Katayama}},
  \bibinfo {author} {\bibfnamefont {T.}~\bibnamefont {Kiguchi}}, \bibinfo
  {author} {\bibfnamefont {A.}~\bibnamefont {Akama}}, \bibinfo {author}
  {\bibfnamefont {T.~J.}\ \bibnamefont {Konno}},\ and\ \bibinfo {author}
  {\bibfnamefont {H.}~\bibnamefont {Funakubo}},\ }\bibfield  {title} {\bibinfo
  {title} {Growth of epitaxial orthorhombic {YO$_{1.5}$}-substituted {HfO$_2$}
  thin film},\ }\href {https://doi.org/10.1063/1.4927450} {\bibfield  {journal}
  {\bibinfo  {journal} {Appl. Phys. Lett.}\ }\textbf {\bibinfo {volume}
  {107}},\ \bibinfo {pages} {032910} (\bibinfo {year} {2015})}\BibitemShut
  {NoStop}%
\bibitem [{\citenamefont {Katayama}\ \emph
  {et~al.}(2016{\natexlab{a}})\citenamefont {Katayama}, \citenamefont
  {Shimizu}, \citenamefont {Sakata}, \citenamefont {Shiraishi}, \citenamefont
  {Nakamura}, \citenamefont {Kiguchi}, \citenamefont {Akama}, \citenamefont
  {Konno}, \citenamefont {Uchida},\ and\ \citenamefont
  {Funakubo}}]{Katayama16p134101}%
  \BibitemOpen
  \bibfield  {author} {\bibinfo {author} {\bibfnamefont {K.}~\bibnamefont
  {Katayama}}, \bibinfo {author} {\bibfnamefont {T.}~\bibnamefont {Shimizu}},
  \bibinfo {author} {\bibfnamefont {O.}~\bibnamefont {Sakata}}, \bibinfo
  {author} {\bibfnamefont {T.}~\bibnamefont {Shiraishi}}, \bibinfo {author}
  {\bibfnamefont {S.}~\bibnamefont {Nakamura}}, \bibinfo {author}
  {\bibfnamefont {T.}~\bibnamefont {Kiguchi}}, \bibinfo {author} {\bibfnamefont
  {A.}~\bibnamefont {Akama}}, \bibinfo {author} {\bibfnamefont {T.~J.}\
  \bibnamefont {Konno}}, \bibinfo {author} {\bibfnamefont {H.}~\bibnamefont
  {Uchida}},\ and\ \bibinfo {author} {\bibfnamefont {H.}~\bibnamefont
  {Funakubo}},\ }\bibfield  {title} {\bibinfo {title} {Orientation control and
  domain structure analysis of \{100\}-oriented epitaxial ferroelectric
  orthorhombic {HfO$_2$}-based thin films},\ }\href
  {https://doi.org/10.1063/1.4945029} {\bibfield  {journal} {\bibinfo
  {journal} {J. Appl. Phys.}\ }\textbf {\bibinfo {volume} {119}},\ \bibinfo
  {pages} {134101} (\bibinfo {year} {2016}{\natexlab{a}})}\BibitemShut
  {NoStop}%
\bibitem [{\citenamefont {Shimizu}\ \emph {et~al.}(2016)\citenamefont
  {Shimizu}, \citenamefont {Katayama}, \citenamefont {Kiguchi}, \citenamefont
  {Akama}, \citenamefont {Konno}, \citenamefont {Sakata},\ and\ \citenamefont
  {Funakubo}}]{Shimizu16p32931}%
  \BibitemOpen
  \bibfield  {author} {\bibinfo {author} {\bibfnamefont {T.}~\bibnamefont
  {Shimizu}}, \bibinfo {author} {\bibfnamefont {K.}~\bibnamefont {Katayama}},
  \bibinfo {author} {\bibfnamefont {T.}~\bibnamefont {Kiguchi}}, \bibinfo
  {author} {\bibfnamefont {A.}~\bibnamefont {Akama}}, \bibinfo {author}
  {\bibfnamefont {T.~J.}\ \bibnamefont {Konno}}, \bibinfo {author}
  {\bibfnamefont {O.}~\bibnamefont {Sakata}},\ and\ \bibinfo {author}
  {\bibfnamefont {H.}~\bibnamefont {Funakubo}},\ }\bibfield  {title} {\bibinfo
  {title} {The demonstration of significant ferroelectricity in epitaxial
  {Y}-doped {HfO$_2$} film},\ }\href {https://doi.org/10.1038/srep32931}
  {\bibfield  {journal} {\bibinfo  {journal} {Sci. Rep.}\ }\textbf {\bibinfo
  {volume} {6}},\ \bibinfo {pages} {32931} (\bibinfo {year}
  {2016})}\BibitemShut {NoStop}%
\bibitem [{\citenamefont {Katayama}\ \emph
  {et~al.}(2016{\natexlab{b}})\citenamefont {Katayama}, \citenamefont
  {Shimizu}, \citenamefont {Sakata}, \citenamefont {Shiraishi}, \citenamefont
  {Nakamura}, \citenamefont {Kiguchi}, \citenamefont {Akama}, \citenamefont
  {Konno}, \citenamefont {Uchida},\ and\ \citenamefont
  {Funakubo}}]{Katayama16p112901}%
  \BibitemOpen
  \bibfield  {author} {\bibinfo {author} {\bibfnamefont {K.}~\bibnamefont
  {Katayama}}, \bibinfo {author} {\bibfnamefont {T.}~\bibnamefont {Shimizu}},
  \bibinfo {author} {\bibfnamefont {O.}~\bibnamefont {Sakata}}, \bibinfo
  {author} {\bibfnamefont {T.}~\bibnamefont {Shiraishi}}, \bibinfo {author}
  {\bibfnamefont {S.}~\bibnamefont {Nakamura}}, \bibinfo {author}
  {\bibfnamefont {T.}~\bibnamefont {Kiguchi}}, \bibinfo {author} {\bibfnamefont
  {A.}~\bibnamefont {Akama}}, \bibinfo {author} {\bibfnamefont {T.~J.}\
  \bibnamefont {Konno}}, \bibinfo {author} {\bibfnamefont {H.}~\bibnamefont
  {Uchida}},\ and\ \bibinfo {author} {\bibfnamefont {H.}~\bibnamefont
  {Funakubo}},\ }\bibfield  {title} {\bibinfo {title} {Growth of (111)-oriented
  epitaxial and textured ferroelectric {Y}-doped {HfO$_2$} films for downscaled
  devices},\ }\href {https://doi.org/10.1063/1.4962431} {\bibfield  {journal}
  {\bibinfo  {journal} {Appl. Phys. Lett.}\ }\textbf {\bibinfo {volume}
  {109}},\ \bibinfo {pages} {112901} (\bibinfo {year}
  {2016}{\natexlab{b}})}\BibitemShut {NoStop}%
\bibitem [{\citenamefont {Mimura}\ \emph {et~al.}(2019)\citenamefont {Mimura},
  \citenamefont {Shimizu}, \citenamefont {Kiguchi}, \citenamefont {Akama},
  \citenamefont {Konno}, \citenamefont {Katsuya}, \citenamefont {Sakata},\ and\
  \citenamefont {Funakubo}}]{Mimura19pSBBB09}%
  \BibitemOpen
  \bibfield  {author} {\bibinfo {author} {\bibfnamefont {T.}~\bibnamefont
  {Mimura}}, \bibinfo {author} {\bibfnamefont {T.}~\bibnamefont {Shimizu}},
  \bibinfo {author} {\bibfnamefont {T.}~\bibnamefont {Kiguchi}}, \bibinfo
  {author} {\bibfnamefont {A.}~\bibnamefont {Akama}}, \bibinfo {author}
  {\bibfnamefont {T.~J.}\ \bibnamefont {Konno}}, \bibinfo {author}
  {\bibfnamefont {Y.}~\bibnamefont {Katsuya}}, \bibinfo {author} {\bibfnamefont
  {O.}~\bibnamefont {Sakata}},\ and\ \bibinfo {author} {\bibfnamefont
  {H.}~\bibnamefont {Funakubo}},\ }\bibfield  {title} {\bibinfo {title}
  {Effects of heat treatment and in situ high-temperature {X}-ray diffraction
  study on the formation of ferroelectric epitaxial {Y}-doped {HfO$_2$} film},\
  }\href {https://doi.org/10.7567/1347-4065/aafed1} {\bibfield  {journal}
  {\bibinfo  {journal} {Jpn. J. Appl. Phys.}\ }\textbf {\bibinfo {volume}
  {58}},\ \bibinfo {pages} {SBBB09} (\bibinfo {year} {2019})}\BibitemShut
  {NoStop}%
\bibitem [{\citenamefont {Estand{\'i}a}\ \emph {et~al.}(2019)\citenamefont
  {Estand{\'i}a}, \citenamefont {Dix}, \citenamefont {Gazquez}, \citenamefont
  {Fina}, \citenamefont {Lyu}, \citenamefont {Chisholm}, \citenamefont
  {Fontcuberta},\ and\ \citenamefont {S{\'a}nchez}}]{Estandia19p1449}%
  \BibitemOpen
  \bibfield  {author} {\bibinfo {author} {\bibfnamefont {S.}~\bibnamefont
  {Estand{\'i}a}}, \bibinfo {author} {\bibfnamefont {N.}~\bibnamefont {Dix}},
  \bibinfo {author} {\bibfnamefont {J.}~\bibnamefont {Gazquez}}, \bibinfo
  {author} {\bibfnamefont {I.}~\bibnamefont {Fina}}, \bibinfo {author}
  {\bibfnamefont {J.}~\bibnamefont {Lyu}}, \bibinfo {author} {\bibfnamefont
  {M.~F.}\ \bibnamefont {Chisholm}}, \bibinfo {author} {\bibfnamefont
  {J.}~\bibnamefont {Fontcuberta}},\ and\ \bibinfo {author} {\bibfnamefont
  {F.}~\bibnamefont {S{\'a}nchez}},\ }\bibfield  {title} {\bibinfo {title}
  {Engineering ferroelectric {Hf$_{0.5}$Zr$_{0.5}$O$_2$} thin films by
  epitaxial stress},\ }\href {https://doi.org/10.1021/acsaelm.9b00256}
  {\bibfield  {journal} {\bibinfo  {journal} {ACS Appl. Electron. Mater.}\
  }\textbf {\bibinfo {volume} {1}},\ \bibinfo {pages} {1449} (\bibinfo {year}
  {2019})}\BibitemShut {NoStop}%
\bibitem [{\citenamefont {Yun}\ \emph {et~al.}(2022)\citenamefont {Yun},
  \citenamefont {Buragohain}, \citenamefont {Li}, \citenamefont {Ahmadi},
  \citenamefont {Zhang}, \citenamefont {Li}, \citenamefont {Wang},
  \citenamefont {Li}, \citenamefont {Lu}, \citenamefont {Tao}, \citenamefont
  {Wang}, \citenamefont {Shield}, \citenamefont {Tsymbal}, \citenamefont
  {Gruverman},\ and\ \citenamefont {Xu}}]{Yun22p903}%
  \BibitemOpen
  \bibfield  {author} {\bibinfo {author} {\bibfnamefont {Y.}~\bibnamefont
  {Yun}}, \bibinfo {author} {\bibfnamefont {P.}~\bibnamefont {Buragohain}},
  \bibinfo {author} {\bibfnamefont {M.}~\bibnamefont {Li}}, \bibinfo {author}
  {\bibfnamefont {Z.}~\bibnamefont {Ahmadi}}, \bibinfo {author} {\bibfnamefont
  {Y.}~\bibnamefont {Zhang}}, \bibinfo {author} {\bibfnamefont
  {X.}~\bibnamefont {Li}}, \bibinfo {author} {\bibfnamefont {H.}~\bibnamefont
  {Wang}}, \bibinfo {author} {\bibfnamefont {J.}~\bibnamefont {Li}}, \bibinfo
  {author} {\bibfnamefont {P.}~\bibnamefont {Lu}}, \bibinfo {author}
  {\bibfnamefont {L.}~\bibnamefont {Tao}}, \bibinfo {author} {\bibfnamefont
  {H.}~\bibnamefont {Wang}}, \bibinfo {author} {\bibfnamefont {J.~E.}\
  \bibnamefont {Shield}}, \bibinfo {author} {\bibfnamefont {E.~Y.}\
  \bibnamefont {Tsymbal}}, \bibinfo {author} {\bibfnamefont {A.}~\bibnamefont
  {Gruverman}},\ and\ \bibinfo {author} {\bibfnamefont {X.}~\bibnamefont
  {Xu}},\ }\bibfield  {title} {\bibinfo {title} {Intrinsic ferroelectricity in
  {Y}-doped {HfO$_2$} thin films},\ }\href
  {https://doi.org/10.1038/s41563-022-01282-6} {\bibfield  {journal} {\bibinfo
  {journal} {Nat. Mater.}\ }\textbf {\bibinfo {volume} {21}},\ \bibinfo {pages}
  {903} (\bibinfo {year} {2022})}\BibitemShut {NoStop}%
\bibitem [{\citenamefont {Wei}\ \emph {et~al.}(2018)\citenamefont {Wei},
  \citenamefont {Nukala}, \citenamefont {Salverda}, \citenamefont {Matzen},
  \citenamefont {Zhao}, \citenamefont {Momand}, \citenamefont {Everhardt},
  \citenamefont {Agnus}, \citenamefont {Blake}, \citenamefont {Lecoeur},
  \citenamefont {Kooi}, \citenamefont {{\'I}{\~n}iguez}, \citenamefont
  {Dkhil},\ and\ \citenamefont {Noheda}}]{Wei18p1095}%
  \BibitemOpen
  \bibfield  {author} {\bibinfo {author} {\bibfnamefont {Y.}~\bibnamefont
  {Wei}}, \bibinfo {author} {\bibfnamefont {P.}~\bibnamefont {Nukala}},
  \bibinfo {author} {\bibfnamefont {M.}~\bibnamefont {Salverda}}, \bibinfo
  {author} {\bibfnamefont {S.}~\bibnamefont {Matzen}}, \bibinfo {author}
  {\bibfnamefont {H.~J.}\ \bibnamefont {Zhao}}, \bibinfo {author}
  {\bibfnamefont {J.}~\bibnamefont {Momand}}, \bibinfo {author} {\bibfnamefont
  {A.~S.}\ \bibnamefont {Everhardt}}, \bibinfo {author} {\bibfnamefont
  {G.}~\bibnamefont {Agnus}}, \bibinfo {author} {\bibfnamefont {G.~R.}\
  \bibnamefont {Blake}}, \bibinfo {author} {\bibfnamefont {P.}~\bibnamefont
  {Lecoeur}}, \bibinfo {author} {\bibfnamefont {B.~J.}\ \bibnamefont {Kooi}},
  \bibinfo {author} {\bibfnamefont {J.}~\bibnamefont {{\'I}{\~n}iguez}},
  \bibinfo {author} {\bibfnamefont {B.}~\bibnamefont {Dkhil}},\ and\ \bibinfo
  {author} {\bibfnamefont {B.}~\bibnamefont {Noheda}},\ }\bibfield  {title}
  {\bibinfo {title} {A rhombohedral ferroelectric phase in epitaxially strained
  {Hf$_{0.5}$Zr$_{0.5}$O$_2$} thin films},\ }\href
  {https://doi.org/10.1038/s41563-018-0196-0} {\bibfield  {journal} {\bibinfo
  {journal} {Nat. Mater.}\ }\textbf {\bibinfo {volume} {17}},\ \bibinfo {pages}
  {1095} (\bibinfo {year} {2018})}\BibitemShut {NoStop}%
\bibitem [{\citenamefont {{B{\'e}gon-Lours}}\ \emph {et~al.}(2020)\citenamefont
  {{B{\'e}gon-Lours}}, \citenamefont {Mulder}, \citenamefont {Nukala},
  \citenamefont {{de Graaf}}, \citenamefont {Birkh{\"o}lzer}, \citenamefont
  {Kooi}, \citenamefont {Noheda}, \citenamefont {Koster},\ and\ \citenamefont
  {Rijnders}}]{Begon-Lours20p043401}%
  \BibitemOpen
  \bibfield  {author} {\bibinfo {author} {\bibfnamefont {L.}~\bibnamefont
  {{B{\'e}gon-Lours}}}, \bibinfo {author} {\bibfnamefont {M.}~\bibnamefont
  {Mulder}}, \bibinfo {author} {\bibfnamefont {P.}~\bibnamefont {Nukala}},
  \bibinfo {author} {\bibfnamefont {S.}~\bibnamefont {{de Graaf}}}, \bibinfo
  {author} {\bibfnamefont {Y.~A.}\ \bibnamefont {Birkh{\"o}lzer}}, \bibinfo
  {author} {\bibfnamefont {B.}~\bibnamefont {Kooi}}, \bibinfo {author}
  {\bibfnamefont {B.}~\bibnamefont {Noheda}}, \bibinfo {author} {\bibfnamefont
  {G.}~\bibnamefont {Koster}},\ and\ \bibinfo {author} {\bibfnamefont
  {G.}~\bibnamefont {Rijnders}},\ }\bibfield  {title} {\bibinfo {title}
  {Stabilization of phase-pure rhombohedral {HfZrO$_4$} in pulsed laser
  deposited thin films},\ }\href
  {https://doi.org/10.1103/PhysRevMaterials.4.043401} {\bibfield  {journal}
  {\bibinfo  {journal} {Phys. Rev. Materials}\ }\textbf {\bibinfo {volume}
  {4}},\ \bibinfo {pages} {043401} (\bibinfo {year} {2020})}\BibitemShut
  {NoStop}%
\bibitem [{\citenamefont {Zheng}\ \emph {et~al.}(2021)\citenamefont {Zheng},
  \citenamefont {Yin}, \citenamefont {Cheng}, \citenamefont {Zhang},
  \citenamefont {Wu},\ and\ \citenamefont {Qi}}]{Zheng21p172904}%
  \BibitemOpen
  \bibfield  {author} {\bibinfo {author} {\bibfnamefont {M.}~\bibnamefont
  {Zheng}}, \bibinfo {author} {\bibfnamefont {Z.}~\bibnamefont {Yin}}, \bibinfo
  {author} {\bibfnamefont {Y.}~\bibnamefont {Cheng}}, \bibinfo {author}
  {\bibfnamefont {X.}~\bibnamefont {Zhang}}, \bibinfo {author} {\bibfnamefont
  {J.}~\bibnamefont {Wu}},\ and\ \bibinfo {author} {\bibfnamefont
  {J.}~\bibnamefont {Qi}},\ }\bibfield  {title} {\bibinfo {title}
  {Stabilization of thick, rhombohedral {Hf$_{0.5}$Zr$_{0.5}$O$_2$} epilayer on
  c-plane {ZnO}},\ }\href {https://doi.org/10.1063/5.0064113} {\bibfield
  {journal} {\bibinfo  {journal} {Appl. Phys. Lett.}\ }\textbf {\bibinfo
  {volume} {119}},\ \bibinfo {pages} {172904} (\bibinfo {year}
  {2021})}\BibitemShut {NoStop}%
\bibitem [{\citenamefont {Huan}\ \emph {et~al.}(2014)\citenamefont {Huan},
  \citenamefont {Sharma}, \citenamefont {Rossetti},\ and\ \citenamefont
  {Ramprasad}}]{Huan14p064111}%
  \BibitemOpen
  \bibfield  {author} {\bibinfo {author} {\bibfnamefont {T.~D.}\ \bibnamefont
  {Huan}}, \bibinfo {author} {\bibfnamefont {V.}~\bibnamefont {Sharma}},
  \bibinfo {author} {\bibfnamefont {G.~A.}\ \bibnamefont {Rossetti}},\ and\
  \bibinfo {author} {\bibfnamefont {R.}~\bibnamefont {Ramprasad}},\ }\bibfield
  {title} {\bibinfo {title} {Pathways towards ferroelectricity in hafnia},\
  }\href {https://doi.org/10.1103/PhysRevB.90.064111} {\bibfield  {journal}
  {\bibinfo  {journal} {Phys. Rev. B}\ }\textbf {\bibinfo {volume} {90}},\
  \bibinfo {pages} {064111} (\bibinfo {year} {2014})}\BibitemShut {NoStop}%
\bibitem [{\citenamefont {Qi}\ \emph {et~al.}(2020)\citenamefont {Qi},
  \citenamefont {Singh}, \citenamefont {Lau}, \citenamefont {Huang},
  \citenamefont {Xu}, \citenamefont {Walker}, \citenamefont {Ahn},
  \citenamefont {Cheong},\ and\ \citenamefont {Rabe}}]{Qi20p257603}%
  \BibitemOpen
  \bibfield  {author} {\bibinfo {author} {\bibfnamefont {Y.}~\bibnamefont
  {Qi}}, \bibinfo {author} {\bibfnamefont {S.}~\bibnamefont {Singh}}, \bibinfo
  {author} {\bibfnamefont {C.}~\bibnamefont {Lau}}, \bibinfo {author}
  {\bibfnamefont {F.-T.}\ \bibnamefont {Huang}}, \bibinfo {author}
  {\bibfnamefont {X.}~\bibnamefont {Xu}}, \bibinfo {author} {\bibfnamefont
  {F.~J.}\ \bibnamefont {Walker}}, \bibinfo {author} {\bibfnamefont {C.~H.}\
  \bibnamefont {Ahn}}, \bibinfo {author} {\bibfnamefont {S.-W.}\ \bibnamefont
  {Cheong}},\ and\ \bibinfo {author} {\bibfnamefont {K.~M.}\ \bibnamefont
  {Rabe}},\ }\bibfield  {title} {\bibinfo {title} {Stabilization of competing
  ferroelectric phases of {HfO$_2$} under epitaxial strain},\ }\href
  {https://doi.org/10.1103/PhysRevLett.125.257603} {\bibfield  {journal}
  {\bibinfo  {journal} {Phys. Rev. Lett.}\ }\textbf {\bibinfo {volume} {125}},\
  \bibinfo {pages} {257603} (\bibinfo {year} {2020})}\BibitemShut {NoStop}%
\bibitem [{\citenamefont {Materlik}\ \emph {et~al.}(2015)\citenamefont
  {Materlik}, \citenamefont {K{\"u}nneth},\ and\ \citenamefont
  {Kersch}}]{Materlik15p134109}%
  \BibitemOpen
  \bibfield  {author} {\bibinfo {author} {\bibfnamefont {R.}~\bibnamefont
  {Materlik}}, \bibinfo {author} {\bibfnamefont {C.}~\bibnamefont
  {K{\"u}nneth}},\ and\ \bibinfo {author} {\bibfnamefont {A.}~\bibnamefont
  {Kersch}},\ }\bibfield  {title} {\bibinfo {title} {The origin of
  ferroelectricity in {Hf$_{1-x}$Zr$_x$O$_2$}: A computational investigation
  and a surface energy model},\ }\href {https://doi.org/10.1063/1.4916707}
  {\bibfield  {journal} {\bibinfo  {journal} {J. Appl. Phys.}\ }\textbf
  {\bibinfo {volume} {117}},\ \bibinfo {pages} {134109} (\bibinfo {year}
  {2015})}\BibitemShut {NoStop}%
\bibitem [{\citenamefont {Batra}\ \emph {et~al.}(2016)\citenamefont {Batra},
  \citenamefont {Tran},\ and\ \citenamefont {Ramprasad}}]{Batra16p172902}%
  \BibitemOpen
  \bibfield  {author} {\bibinfo {author} {\bibfnamefont {R.}~\bibnamefont
  {Batra}}, \bibinfo {author} {\bibfnamefont {H.~D.}\ \bibnamefont {Tran}},\
  and\ \bibinfo {author} {\bibfnamefont {R.}~\bibnamefont {Ramprasad}},\
  }\bibfield  {title} {\bibinfo {title} {Stabilization of metastable phases in
  hafnia owing to surface energy effects},\ }\href
  {https://doi.org/10.1063/1.4947490} {\bibfield  {journal} {\bibinfo
  {journal} {Appl. Phys. Lett.}\ }\textbf {\bibinfo {volume} {108}},\ \bibinfo
  {pages} {172902} (\bibinfo {year} {2016})}\BibitemShut {NoStop}%
\bibitem [{\citenamefont {Wu}\ \emph {et~al.}(2020)\citenamefont {Wu},
  \citenamefont {Mo}, \citenamefont {Saraya}, \citenamefont {Hiramoto},\ and\
  \citenamefont {Kobayashi}}]{Wu20p252904}%
  \BibitemOpen
  \bibfield  {author} {\bibinfo {author} {\bibfnamefont {J.}~\bibnamefont
  {Wu}}, \bibinfo {author} {\bibfnamefont {F.}~\bibnamefont {Mo}}, \bibinfo
  {author} {\bibfnamefont {T.}~\bibnamefont {Saraya}}, \bibinfo {author}
  {\bibfnamefont {T.}~\bibnamefont {Hiramoto}},\ and\ \bibinfo {author}
  {\bibfnamefont {M.}~\bibnamefont {Kobayashi}},\ }\bibfield  {title} {\bibinfo
  {title} {A first-principles study on ferroelectric phase formation of
  {Si}-doped {HfO$_2$} through nucleation and phase transition in thermal
  process},\ }\href {https://doi.org/10.1063/5.0035139} {\bibfield  {journal}
  {\bibinfo  {journal} {Appl. Phys. Lett.}\ }\textbf {\bibinfo {volume}
  {117}},\ \bibinfo {pages} {252904} (\bibinfo {year} {2020})}\BibitemShut
  {NoStop}%
\bibitem [{\citenamefont {Batra}\ \emph {et~al.}(2017)\citenamefont {Batra},
  \citenamefont {Huan}, \citenamefont {Rossetti},\ and\ \citenamefont
  {Ramprasad}}]{Batra17p9102}%
  \BibitemOpen
  \bibfield  {author} {\bibinfo {author} {\bibfnamefont {R.}~\bibnamefont
  {Batra}}, \bibinfo {author} {\bibfnamefont {T.~D.}\ \bibnamefont {Huan}},
  \bibinfo {author} {\bibfnamefont {G.~A.~J.}\ \bibnamefont {Rossetti}},\ and\
  \bibinfo {author} {\bibfnamefont {R.}~\bibnamefont {Ramprasad}},\ }\bibfield
  {title} {\bibinfo {title} {Dopants promoting ferroelectricity in hafnia:
  Insights from a comprehensive chemical space exploration},\ }\href
  {https://doi.org/10.1021/acs.chemmater.7b02835} {\bibfield  {journal}
  {\bibinfo  {journal} {Chem. Mater.}\ }\textbf {\bibinfo {volume} {29}},\
  \bibinfo {pages} {9102} (\bibinfo {year} {2017})}\BibitemShut {NoStop}%
\bibitem [{\citenamefont {Materlik}\ \emph {et~al.}(2018)\citenamefont
  {Materlik}, \citenamefont {K{\"u}nneth}, \citenamefont {Falkowski},
  \citenamefont {Mikolajick},\ and\ \citenamefont
  {Kersch}}]{Materlik18p164101}%
  \BibitemOpen
  \bibfield  {author} {\bibinfo {author} {\bibfnamefont {R.}~\bibnamefont
  {Materlik}}, \bibinfo {author} {\bibfnamefont {C.}~\bibnamefont
  {K{\"u}nneth}}, \bibinfo {author} {\bibfnamefont {M.}~\bibnamefont
  {Falkowski}}, \bibinfo {author} {\bibfnamefont {T.}~\bibnamefont
  {Mikolajick}},\ and\ \bibinfo {author} {\bibfnamefont {A.}~\bibnamefont
  {Kersch}},\ }\bibfield  {title} {\bibinfo {title} {{Al}-, {Y}-, and
  {La}-doping effects favoring intrinsic and field induced ferroelectricity in
  {HfO$_2$}: A first principles study},\ }\href
  {https://doi.org/10.1063/1.5021746} {\bibfield  {journal} {\bibinfo
  {journal} {J. Appl. Phys.}\ }\textbf {\bibinfo {volume} {123}},\ \bibinfo
  {pages} {164101} (\bibinfo {year} {2018})}\BibitemShut {NoStop}%
\bibitem [{\citenamefont {Ma}\ and\ \citenamefont {Liu}(2023)}]{Ma23p096801}%
  \BibitemOpen
  \bibfield  {author} {\bibinfo {author} {\bibfnamefont {L.-Y.}\ \bibnamefont
  {Ma}}\ and\ \bibinfo {author} {\bibfnamefont {S.}~\bibnamefont {Liu}},\
  }\bibfield  {title} {\bibinfo {title} {Structural polymorphism kinetics
  promoted by charged oxygen vacancies in {HfO$_2$}},\ }\href
  {https://doi.org/10.1103/PhysRevLett.130.096801} {\bibfield  {journal}
  {\bibinfo  {journal} {Phys. Rev. Lett.}\ }\textbf {\bibinfo {volume} {130}},\
  \bibinfo {pages} {096801} (\bibinfo {year} {2023})}\BibitemShut {NoStop}%
\bibitem [{\citenamefont {Schroeder}\ \emph {et~al.}(2022)\citenamefont
  {Schroeder}, \citenamefont {Park}, \citenamefont {Mikolajick},\ and\
  \citenamefont {Hwang}}]{Schroeder22p653}%
  \BibitemOpen
  \bibfield  {author} {\bibinfo {author} {\bibfnamefont {U.}~\bibnamefont
  {Schroeder}}, \bibinfo {author} {\bibfnamefont {M.~H.}\ \bibnamefont {Park}},
  \bibinfo {author} {\bibfnamefont {T.}~\bibnamefont {Mikolajick}},\ and\
  \bibinfo {author} {\bibfnamefont {C.~S.}\ \bibnamefont {Hwang}},\ }\bibfield
  {title} {\bibinfo {title} {The fundamentals and applications of ferroelectric
  {HfO$_2$}},\ }\href {https://doi.org/10.1038/s41578-022-00431-2} {\bibfield
  {journal} {\bibinfo  {journal} {Nat. Rev. Mater.}\ }\textbf {\bibinfo
  {volume} {7}},\ \bibinfo {pages} {653} (\bibinfo {year} {2022})}\BibitemShut
  {NoStop}%
\bibitem [{\citenamefont {Paull}\ \emph {et~al.}(2022)\citenamefont {Paull},
  \citenamefont {Xu}, \citenamefont {Cheng}, \citenamefont {Zhang},
  \citenamefont {Xu}, \citenamefont {Kelley}, \citenamefont {{de Marco}},
  \citenamefont {Vasudevan}, \citenamefont {Bellaiche}, \citenamefont
  {Nagarajan},\ and\ \citenamefont {Sando}}]{Paull22p74}%
  \BibitemOpen
  \bibfield  {author} {\bibinfo {author} {\bibfnamefont {O.}~\bibnamefont
  {Paull}}, \bibinfo {author} {\bibfnamefont {C.}~\bibnamefont {Xu}}, \bibinfo
  {author} {\bibfnamefont {X.}~\bibnamefont {Cheng}}, \bibinfo {author}
  {\bibfnamefont {Y.}~\bibnamefont {Zhang}}, \bibinfo {author} {\bibfnamefont
  {B.}~\bibnamefont {Xu}}, \bibinfo {author} {\bibfnamefont {K.~P.}\
  \bibnamefont {Kelley}}, \bibinfo {author} {\bibfnamefont {A.}~\bibnamefont
  {{de Marco}}}, \bibinfo {author} {\bibfnamefont {R.~K.}\ \bibnamefont
  {Vasudevan}}, \bibinfo {author} {\bibfnamefont {L.}~\bibnamefont
  {Bellaiche}}, \bibinfo {author} {\bibfnamefont {V.}~\bibnamefont
  {Nagarajan}},\ and\ \bibinfo {author} {\bibfnamefont {D.}~\bibnamefont
  {Sando}},\ }\bibfield  {title} {\bibinfo {title} {Anisotropic epitaxial
  stabilization of a low-symmetry ferroelectric with enhanced electromechanical
  response},\ }\href {https://doi.org/10.1038/s41563-021-01098-w} {\bibfield
  {journal} {\bibinfo  {journal} {Nat. Mater.}\ }\textbf {\bibinfo {volume}
  {21}},\ \bibinfo {pages} {74} (\bibinfo {year} {2022})}\BibitemShut {NoStop}%
\bibitem [{\citenamefont {Liu}\ \emph {et~al.}(2023)\citenamefont {Liu},
  \citenamefont {Jin}, \citenamefont {Zhang}, \citenamefont {Liu},
  \citenamefont {Zhang}, \citenamefont {Hua}, \citenamefont {Zhang},
  \citenamefont {Ye}, \citenamefont {Gao}, \citenamefont {Ma}, \citenamefont
  {Wang},\ and\ \citenamefont {Wu}}]{Liu23p2209925}%
  \BibitemOpen
  \bibfield  {author} {\bibinfo {author} {\bibfnamefont {K.}~\bibnamefont
  {Liu}}, \bibinfo {author} {\bibfnamefont {F.}~\bibnamefont {Jin}}, \bibinfo
  {author} {\bibfnamefont {X.}~\bibnamefont {Zhang}}, \bibinfo {author}
  {\bibfnamefont {K.}~\bibnamefont {Liu}}, \bibinfo {author} {\bibfnamefont
  {Z.}~\bibnamefont {Zhang}}, \bibinfo {author} {\bibfnamefont
  {E.}~\bibnamefont {Hua}}, \bibinfo {author} {\bibfnamefont {J.}~\bibnamefont
  {Zhang}}, \bibinfo {author} {\bibfnamefont {H.}~\bibnamefont {Ye}}, \bibinfo
  {author} {\bibfnamefont {G.}~\bibnamefont {Gao}}, \bibinfo {author}
  {\bibfnamefont {C.}~\bibnamefont {Ma}}, \bibinfo {author} {\bibfnamefont
  {L.}~\bibnamefont {Wang}},\ and\ \bibinfo {author} {\bibfnamefont
  {W.}~\bibnamefont {Wu}},\ }\bibfield  {title} {\bibinfo {title} {Anisotropic
  strain-mediated symmetry engineering and enhancement of ferroelectricity in
  {Hf$_{0.5}$Zr$_{0.5}$O$_2$}/{La$_{0.67}$Sr$_{0.33}$MnO$_3$}
  heterostructures},\ }\href {https://doi.org/10.1002/adfm.202209925}
  {\bibfield  {journal} {\bibinfo  {journal} {Adv. Funct. Mater.}\ }\textbf
  {\bibinfo {volume} {33}},\ \bibinfo {pages} {2209925} (\bibinfo {year}
  {2023})}\BibitemShut {NoStop}%
\bibitem [{\citenamefont {Liu}\ and\ \citenamefont
  {Hanrahan}(2019)}]{Liu19p054404}%
  \BibitemOpen
  \bibfield  {author} {\bibinfo {author} {\bibfnamefont {S.}~\bibnamefont
  {Liu}}\ and\ \bibinfo {author} {\bibfnamefont {B.~M.}\ \bibnamefont
  {Hanrahan}},\ }\bibfield  {title} {\bibinfo {title} {Effects of growth
  orientations and epitaxial strains on phase stability of {HfO$_2$} thin
  films},\ }\href {https://doi.org/10.1103/PhysRevMaterials.3.054404}
  {\bibfield  {journal} {\bibinfo  {journal} {Phys. Rev. Materials}\ }\textbf
  {\bibinfo {volume} {3}},\ \bibinfo {pages} {054404} (\bibinfo {year}
  {2019})}\BibitemShut {NoStop}%
\bibitem [{\citenamefont {Chae}\ \emph {et~al.}(2020)\citenamefont {Chae},
  \citenamefont {Hwang}, \citenamefont {Chagarov}, \citenamefont {Kummel},\
  and\ \citenamefont {Cho}}]{Chae20p054101}%
  \BibitemOpen
  \bibfield  {author} {\bibinfo {author} {\bibfnamefont {K.}~\bibnamefont
  {Chae}}, \bibinfo {author} {\bibfnamefont {J.}~\bibnamefont {Hwang}},
  \bibinfo {author} {\bibfnamefont {E.}~\bibnamefont {Chagarov}}, \bibinfo
  {author} {\bibfnamefont {A.}~\bibnamefont {Kummel}},\ and\ \bibinfo {author}
  {\bibfnamefont {K.}~\bibnamefont {Cho}},\ }\bibfield  {title} {\bibinfo
  {title} {Stability of ferroelectric and antiferroelectric hafnium-zirconium
  oxide thin films},\ }\href {https://doi.org/10.1063/5.0011547} {\bibfield
  {journal} {\bibinfo  {journal} {J. Appl. Phys.}\ }\textbf {\bibinfo {volume}
  {128}},\ \bibinfo {pages} {054101} (\bibinfo {year} {2020})}\BibitemShut
  {NoStop}%
\bibitem [{\citenamefont {Torrej{\'o}n}\ \emph {et~al.}(2018)\citenamefont
  {Torrej{\'o}n}, \citenamefont {Langenberg}, \citenamefont {Mag{\'e}n},
  \citenamefont {Larrea}, \citenamefont {Blasco}, \citenamefont {Santiso},
  \citenamefont {Algarabel},\ and\ \citenamefont {Pardo}}]{Torrejon18p013401}%
  \BibitemOpen
  \bibfield  {author} {\bibinfo {author} {\bibfnamefont {L.}~\bibnamefont
  {Torrej{\'o}n}}, \bibinfo {author} {\bibfnamefont {E.}~\bibnamefont
  {Langenberg}}, \bibinfo {author} {\bibfnamefont {C.}~\bibnamefont
  {Mag{\'e}n}}, \bibinfo {author} {\bibfnamefont {{\'A}.}~\bibnamefont
  {Larrea}}, \bibinfo {author} {\bibfnamefont {J.}~\bibnamefont {Blasco}},
  \bibinfo {author} {\bibfnamefont {J.}~\bibnamefont {Santiso}}, \bibinfo
  {author} {\bibfnamefont {P.~A.}\ \bibnamefont {Algarabel}},\ and\ \bibinfo
  {author} {\bibfnamefont {J.~A.}\ \bibnamefont {Pardo}},\ }\bibfield  {title}
  {\bibinfo {title} {Growth and structural characterization of strained
  epitaxial {Hf$_{0.5}$Zr$_{0.5}$O$_2$} thin films},\ }\href
  {https://doi.org/10.1103/PhysRevMaterials.2.013401} {\bibfield  {journal}
  {\bibinfo  {journal} {Phys. Rev. Materials}\ }\textbf {\bibinfo {volume}
  {2}},\ \bibinfo {pages} {013401} (\bibinfo {year} {2018})}\BibitemShut
  {NoStop}%
\bibitem [{\citenamefont {Fina}\ and\ \citenamefont
  {S{\'a}nchez}(2021)}]{Fina21p1530}%
  \BibitemOpen
  \bibfield  {author} {\bibinfo {author} {\bibfnamefont {I.}~\bibnamefont
  {Fina}}\ and\ \bibinfo {author} {\bibfnamefont {F.}~\bibnamefont
  {S{\'a}nchez}},\ }\bibfield  {title} {\bibinfo {title} {Epitaxial
  ferroelectric {HfO$_2$} films: Growth, properties, and devices},\ }\href
  {https://doi.org/10.1021/acsaelm.1c00110} {\bibfield  {journal} {\bibinfo
  {journal} {ACS Appl. Electron. Mater.}\ }\textbf {\bibinfo {volume} {3}},\
  \bibinfo {pages} {1530} (\bibinfo {year} {2021})}\BibitemShut {NoStop}%
\bibitem [{\citenamefont {Zhang}\ \emph {et~al.}(2020)\citenamefont {Zhang},
  \citenamefont {Yang}, \citenamefont {Tao}, \citenamefont {Tsymbal},\ and\
  \citenamefont {Alexandrov}}]{Zhang20p014068}%
  \BibitemOpen
  \bibfield  {author} {\bibinfo {author} {\bibfnamefont {Y.}~\bibnamefont
  {Zhang}}, \bibinfo {author} {\bibfnamefont {Q.}~\bibnamefont {Yang}},
  \bibinfo {author} {\bibfnamefont {L.}~\bibnamefont {Tao}}, \bibinfo {author}
  {\bibfnamefont {E.~Y.}\ \bibnamefont {Tsymbal}},\ and\ \bibinfo {author}
  {\bibfnamefont {V.}~\bibnamefont {Alexandrov}},\ }\bibfield  {title}
  {\bibinfo {title} {Effects of strain and film thickness on the stability of
  the rhombohedral phase of {HfO$_2$}},\ }\href
  {https://doi.org/10.1103/PhysRevApplied.14.014068} {\bibfield  {journal}
  {\bibinfo  {journal} {Phys. Rev. Applied}\ }\textbf {\bibinfo {volume}
  {14}},\ \bibinfo {pages} {014068} (\bibinfo {year} {2020})}\BibitemShut
  {NoStop}%
\bibitem [{\citenamefont {Estand{\'i}a}\ \emph {et~al.}(2020)\citenamefont
  {Estand{\'i}a}, \citenamefont {Dix}, \citenamefont {Chisholm}, \citenamefont
  {Fina},\ and\ \citenamefont {S{\'a}nchez}}]{Estandia20p3801}%
  \BibitemOpen
  \bibfield  {author} {\bibinfo {author} {\bibfnamefont {S.}~\bibnamefont
  {Estand{\'i}a}}, \bibinfo {author} {\bibfnamefont {N.}~\bibnamefont {Dix}},
  \bibinfo {author} {\bibfnamefont {M.~F.}\ \bibnamefont {Chisholm}}, \bibinfo
  {author} {\bibfnamefont {I.}~\bibnamefont {Fina}},\ and\ \bibinfo {author}
  {\bibfnamefont {F.}~\bibnamefont {S{\'a}nchez}},\ }\bibfield  {title}
  {\bibinfo {title} {Domain-matching epitaxy of ferroelectric
  {Hf$_{0.5}$Zr$_{0.5}$O$_2$}(111) on {La$_{2/3}$Sr$_{1/3}$MnO$_3$}(001)},\
  }\href {https://doi.org/10.1021/acs.cgd.0c00095} {\bibfield  {journal}
  {\bibinfo  {journal} {Cryst. Growth Des.}\ }\textbf {\bibinfo {volume}
  {20}},\ \bibinfo {pages} {3801} (\bibinfo {year} {2020})}\BibitemShut
  {NoStop}%
\bibitem [{\citenamefont {Jiao}\ \emph {et~al.}(2021)\citenamefont {Jiao},
  \citenamefont {Li}, \citenamefont {Xi}, \citenamefont {Zhang}, \citenamefont
  {Wang}, \citenamefont {Yang}, \citenamefont {Deng},\ and\ \citenamefont
  {Wu}}]{Jiao21p252901}%
  \BibitemOpen
  \bibfield  {author} {\bibinfo {author} {\bibfnamefont {P.}~\bibnamefont
  {Jiao}}, \bibinfo {author} {\bibfnamefont {J.}~\bibnamefont {Li}}, \bibinfo
  {author} {\bibfnamefont {Z.}~\bibnamefont {Xi}}, \bibinfo {author}
  {\bibfnamefont {X.}~\bibnamefont {Zhang}}, \bibinfo {author} {\bibfnamefont
  {J.}~\bibnamefont {Wang}}, \bibinfo {author} {\bibfnamefont {Y.}~\bibnamefont
  {Yang}}, \bibinfo {author} {\bibfnamefont {Y.}~\bibnamefont {Deng}},\ and\
  \bibinfo {author} {\bibfnamefont {D.}~\bibnamefont {Wu}},\ }\bibfield
  {title} {\bibinfo {title} {Ferroelectric {Hf$_{0.5}$Zr$_{0.5}$O$_2$} thin
  films deposited epitaxially on (110)-oriented {SrTiO$_3$}},\ }\href
  {https://doi.org/10.1063/5.0075391} {\bibfield  {journal} {\bibinfo
  {journal} {Appl. Phys. Lett.}\ }\textbf {\bibinfo {volume} {119}},\ \bibinfo
  {pages} {252901} (\bibinfo {year} {2021})}\BibitemShut {NoStop}%
\bibitem [{\citenamefont {Cheema}\ \emph {et~al.}(2020)\citenamefont {Cheema},
  \citenamefont {Kwon}, \citenamefont {Shanker}, \citenamefont {{dos Reis}},
  \citenamefont {Hsu}, \citenamefont {Xiao}, \citenamefont {Zhang},
  \citenamefont {Wagner}, \citenamefont {Datar}, \citenamefont {McCarter},
  \citenamefont {Serrao}, \citenamefont {Yadav}, \citenamefont {Karbasian},
  \citenamefont {Hsu}, \citenamefont {Tan}, \citenamefont {Wang}, \citenamefont
  {Thakare}, \citenamefont {Zhang}, \citenamefont {Mehta}, \citenamefont
  {Karapetrova}, \citenamefont {Chopdekar}, \citenamefont {Shafer},
  \citenamefont {Arenholz}, \citenamefont {Hu}, \citenamefont {Proksch},
  \citenamefont {Ramesh}, \citenamefont {Ciston},\ and\ \citenamefont
  {Salahuddin}}]{Cheema20p478}%
  \BibitemOpen
  \bibfield  {author} {\bibinfo {author} {\bibfnamefont {S.~S.}\ \bibnamefont
  {Cheema}}, \bibinfo {author} {\bibfnamefont {D.}~\bibnamefont {Kwon}},
  \bibinfo {author} {\bibfnamefont {N.}~\bibnamefont {Shanker}}, \bibinfo
  {author} {\bibfnamefont {R.}~\bibnamefont {{dos Reis}}}, \bibinfo {author}
  {\bibfnamefont {S.-L.}\ \bibnamefont {Hsu}}, \bibinfo {author} {\bibfnamefont
  {J.}~\bibnamefont {Xiao}}, \bibinfo {author} {\bibfnamefont {H.}~\bibnamefont
  {Zhang}}, \bibinfo {author} {\bibfnamefont {R.}~\bibnamefont {Wagner}},
  \bibinfo {author} {\bibfnamefont {A.}~\bibnamefont {Datar}}, \bibinfo
  {author} {\bibfnamefont {M.~R.}\ \bibnamefont {McCarter}}, \bibinfo {author}
  {\bibfnamefont {C.~R.}\ \bibnamefont {Serrao}}, \bibinfo {author}
  {\bibfnamefont {A.~K.}\ \bibnamefont {Yadav}}, \bibinfo {author}
  {\bibfnamefont {G.}~\bibnamefont {Karbasian}}, \bibinfo {author}
  {\bibfnamefont {C.-H.}\ \bibnamefont {Hsu}}, \bibinfo {author} {\bibfnamefont
  {A.~J.}\ \bibnamefont {Tan}}, \bibinfo {author} {\bibfnamefont {L.-C.}\
  \bibnamefont {Wang}}, \bibinfo {author} {\bibfnamefont {V.}~\bibnamefont
  {Thakare}}, \bibinfo {author} {\bibfnamefont {X.}~\bibnamefont {Zhang}},
  \bibinfo {author} {\bibfnamefont {A.}~\bibnamefont {Mehta}}, \bibinfo
  {author} {\bibfnamefont {E.}~\bibnamefont {Karapetrova}}, \bibinfo {author}
  {\bibfnamefont {R.~V.}\ \bibnamefont {Chopdekar}}, \bibinfo {author}
  {\bibfnamefont {P.}~\bibnamefont {Shafer}}, \bibinfo {author} {\bibfnamefont
  {E.}~\bibnamefont {Arenholz}}, \bibinfo {author} {\bibfnamefont
  {C.}~\bibnamefont {Hu}}, \bibinfo {author} {\bibfnamefont {R.}~\bibnamefont
  {Proksch}}, \bibinfo {author} {\bibfnamefont {R.}~\bibnamefont {Ramesh}},
  \bibinfo {author} {\bibfnamefont {J.}~\bibnamefont {Ciston}},\ and\ \bibinfo
  {author} {\bibfnamefont {S.}~\bibnamefont {Salahuddin}},\ }\bibfield  {title}
  {\bibinfo {title} {Enhanced ferroelectricity in ultrathin films grown
  directly on silicon},\ }\href {https://doi.org/10.1038/s41586-020-2208-x}
  {\bibfield  {journal} {\bibinfo  {journal} {Nature}\ }\textbf {\bibinfo
  {volume} {580}},\ \bibinfo {pages} {478} (\bibinfo {year}
  {2020})}\BibitemShut {NoStop}%
\bibitem [{\citenamefont {Kresse}\ and\ \citenamefont
  {Furthm{\"u}ller}(1996)}]{Kresse96p11169}%
  \BibitemOpen
  \bibfield  {author} {\bibinfo {author} {\bibfnamefont {G.}~\bibnamefont
  {Kresse}}\ and\ \bibinfo {author} {\bibfnamefont {J.}~\bibnamefont
  {Furthm{\"u}ller}},\ }\bibfield  {title} {\bibinfo {title} {Efficient
  iterative schemes for ab initio total-energy calculations using a plane-wave
  basis set},\ }\href {https://doi.org/10.1103/PhysRevB.54.11169} {\bibfield
  {journal} {\bibinfo  {journal} {Phys. Rev. B}\ }\textbf {\bibinfo {volume}
  {54}},\ \bibinfo {pages} {11169} (\bibinfo {year} {1996})}\BibitemShut
  {NoStop}%
\bibitem [{\citenamefont {Bl{\"o}chl}(1994)}]{Blochl94p17953}%
  \BibitemOpen
  \bibfield  {author} {\bibinfo {author} {\bibfnamefont {P.~E.}\ \bibnamefont
  {Bl{\"o}chl}},\ }\bibfield  {title} {\bibinfo {title} {Projector
  augmented-wave method},\ }\href {https://doi.org/10.1103/PhysRevB.50.17953}
  {\bibfield  {journal} {\bibinfo  {journal} {Phys. Rev. B}\ }\textbf {\bibinfo
  {volume} {50}},\ \bibinfo {pages} {17953} (\bibinfo {year}
  {1994})}\BibitemShut {NoStop}%
\bibitem [{\citenamefont {Kresse}\ and\ \citenamefont
  {Joubert}(1999)}]{Kresse99p1758}%
  \BibitemOpen
  \bibfield  {author} {\bibinfo {author} {\bibfnamefont {G.}~\bibnamefont
  {Kresse}}\ and\ \bibinfo {author} {\bibfnamefont {D.}~\bibnamefont
  {Joubert}},\ }\bibfield  {title} {\bibinfo {title} {From ultrasoft
  pseudopotentials to the projector augmented-wave method},\ }\href
  {https://doi.org/10.1103/PhysRevB.59.1758} {\bibfield  {journal} {\bibinfo
  {journal} {Phys. Rev. B}\ }\textbf {\bibinfo {volume} {59}},\ \bibinfo
  {pages} {1758} (\bibinfo {year} {1999})}\BibitemShut {NoStop}%
\bibitem [{\citenamefont {Perdew}\ \emph {et~al.}(1996)\citenamefont {Perdew},
  \citenamefont {Burke},\ and\ \citenamefont {Ernzerhof}}]{Perdew96p3865}%
  \BibitemOpen
  \bibfield  {author} {\bibinfo {author} {\bibfnamefont {J.~P.}\ \bibnamefont
  {Perdew}}, \bibinfo {author} {\bibfnamefont {K.}~\bibnamefont {Burke}},\ and\
  \bibinfo {author} {\bibfnamefont {M.}~\bibnamefont {Ernzerhof}},\ }\bibfield
  {title} {\bibinfo {title} {Generalized gradient approximation made simple},\
  }\href {https://doi.org/10.1103/PhysRevLett.77.3865} {\bibfield  {journal}
  {\bibinfo  {journal} {Phys. Rev. Lett.}\ }\textbf {\bibinfo {volume} {77}},\
  \bibinfo {pages} {3865} (\bibinfo {year} {1996})}\BibitemShut {NoStop}%
\bibitem [{\citenamefont {Monkhorst}\ and\ \citenamefont
  {Pack}(1976)}]{Monkhorst76p5188}%
  \BibitemOpen
  \bibfield  {author} {\bibinfo {author} {\bibfnamefont {H.~J.}\ \bibnamefont
  {Monkhorst}}\ and\ \bibinfo {author} {\bibfnamefont {J.~D.}\ \bibnamefont
  {Pack}},\ }\bibfield  {title} {\bibinfo {title} {Special points for
  {Brillouin}-zone integrations},\ }\href
  {https://doi.org/10.1103/PhysRevB.13.5188} {\bibfield  {journal} {\bibinfo
  {journal} {Phys. Rev. B}\ }\textbf {\bibinfo {volume} {13}},\ \bibinfo
  {pages} {5188} (\bibinfo {year} {1976})}\BibitemShut {NoStop}%
\bibitem [{\citenamefont {{King-Smith}}\ and\ \citenamefont
  {Vanderbilt}(1993)}]{King-Smith93p1651}%
  \BibitemOpen
  \bibfield  {author} {\bibinfo {author} {\bibfnamefont {R.~D.}\ \bibnamefont
  {{King-Smith}}}\ and\ \bibinfo {author} {\bibfnamefont {D.}~\bibnamefont
  {Vanderbilt}},\ }\bibfield  {title} {\bibinfo {title} {Theory of polarization
  of crystalline solids},\ }\href {https://doi.org/10.1103/PhysRevB.47.1651}
  {\bibfield  {journal} {\bibinfo  {journal} {Phys. Rev. B}\ }\textbf {\bibinfo
  {volume} {47}},\ \bibinfo {pages} {1651} (\bibinfo {year}
  {1993})}\BibitemShut {NoStop}%
\bibitem [{\citenamefont {Vanderbilt}\ and\ \citenamefont
  {{King-Smith}}(1993)}]{Vanderbilt93p4442}%
  \BibitemOpen
  \bibfield  {author} {\bibinfo {author} {\bibfnamefont {D.}~\bibnamefont
  {Vanderbilt}}\ and\ \bibinfo {author} {\bibfnamefont {R.~D.}\ \bibnamefont
  {{King-Smith}}},\ }\bibfield  {title} {\bibinfo {title} {Electric
  polarization as a bulk quantity and its relation to surface charge},\ }\href
  {https://doi.org/10.1103/PhysRevB.48.4442} {\bibfield  {journal} {\bibinfo
  {journal} {Phys. Rev. B}\ }\textbf {\bibinfo {volume} {48}},\ \bibinfo
  {pages} {4442} (\bibinfo {year} {1993})}\BibitemShut {NoStop}%
\end{thebibliography}%

\end{document}